\documentclass[preprintnumbers,amsmath,amssymb,aps,prl,groupedaddress]{revtex4}

\usepackage{multirow}
\usepackage{hyperref}
\usepackage{mathrsfs}
\usepackage{amsmath}
\usepackage{graphicx}
\usepackage{color}
\begin{document}
%\draft
\title{Exceptional entanglement phenomena: non-Hermiticity meeting
non-classicality}
\author{Pei-Rong Han$^{1}$}
\thanks{These authors contribute equally to this work.}
\author{Fan Wu$^{1}$}
\thanks{These authors contribute equally to this work.}
\author{Xin-Jie Huang$^{1}$}
\thanks{These authors contribute equally to this work.}
\author{Huai-Zhi Wu$^{1}$}
\author{Chang-Ling Zou$^{2,3,8}$, Wei Yi$^{2,3,8}$, Mengzhen Zhang$^{4}$, Hekang Li$^{5}$%
	, Kai Xu$^{5,6,8}$, Dongning Zheng$^{5,6,8}$, Heng Fan$^{5,6,8}$}
\author{Jianming Wen$^{7}$}
\thanks{E-mail: jianming.wen@gmail.com}
\author{Zhen-Biao Yang$^{1,8}$}
\thanks{E-mail: zbyang@fzu.edu.cn}
\author{Shi-Biao Zheng$^{1,8}$}
\thanks{E-mail: t96034@fzu.edu.cn}
\address{$^{1}$Fujian Key Laboratory of Quantum Information and Quantum Optics, College of Physics and Information Engineering, Fuzhou University, Fuzhou, Fujian, 350108, China\\
	$^{2}$CAS Key Laboratory of Quantum Information, University of Science and\\
	Technology of China, Hefei 230026, China\\
	$^{3}$CAS Center for Excellence in Quantum Information and Quantum Physics,\\
	University of Science and Technology of China, Hefei 230026, China\\
	$^{4}$Pritzker School of Molecular Engineering, University of Chicago,\\
	Chicago, IL 60637, USA\\
	$^{5}$Institute of Physics, Chinese Academy of Sciences, Beijing 100190,\\
	China\\
	$^{6}$CAS Center for Excellence in Topological Quantum Computation,\\
	University of Chinese Academy of Sciences, Beijing 100190, China\\
	$^{7}$Department of Physics, Kennesaw State University, Marietta, Georgia\\
	30060, USA\\
	$^{8}$Hefei National Laboratory, Hefei 230088, China}
%\date{\today }

\begin{abstract}
Non-Hermitian (NH) extension of quantum-mechanical Hamiltonians represents one of the most significant advancements in physics. During the past two decades,
numerous captivating NH phenomena have been revealed and demonstrated, but
all of which can appear in both quantum and classical systems. This leads to
the fundamental question: What NH signature presents a radical departure
from classical physics? The solution of this problem is indispensable for
exploring genuine NH quantum mechanics, but remains experimentally untouched upon so far.
Here we resolve this basic issue by unveiling distinct exceptional
entanglement phenomena, exemplified by an entanglement transition, occurring
at the exceptional point (EP) of NH interacting quantum systems. We
illustrate and demonstrate such purely quantum-mechanical NH effects with a
naturally-dissipative light-matter system, engineered in a circuit quantum
electrodynamics architecture. Our results lay the foundation for studies of
genuinely quantum-mechanical NH physics, signified by EP-enabled
entanglement behaviors.
\end{abstract}

\vskip0.5cm

\maketitle
\narrowtext

\bigskip When a physical system undergoes dissipation, the Hermiticity of
its Hamiltonian dynamics is broken down. As any system inevitably interacts
with its surrounding environment by exchanging particles or energy,
non-Hermitian (NH) effects are ubiquitous in both classical and quantum
physics. Such effects were once thought to be detrimental, and needed to be
suppressed for observing physical phenomena of interest and for
technological applications, until the discovery that NH effects could
represent a complex extension of quantum mechanics \cite{1,2,3}. Since then,
increasing efforts have been devoted to the exploration of NH physics,
leading to findings of many intriguing phenomena that are uniquely
associated with NH systems. Most of these phenomena are closely related to
the exceptional points (EPs), where both the eigenenergies and the
eigenvectors of the NH Hamiltonian coalesce \cite{4,5,6}. In addition to
fundamental interest, such NH effects promise the realization of enhanced
sensors \cite{7,8,9,10,11}.

Hitherto, there have been a plethora of experimental investigations on
genuinely NH phenomena, ranging from real-to-complex spectral transition to
NH topology \cite{12,13,14,15,16,17,18,19,20}, as well as on relevant applications \cite{21,22,23,24,25}, most of
which were performed with classically interacting but non-entangled systems.
The past few years have witnessed a number of demonstrations of similar NH
phenomena in different quantum systems, ranging from photons \cite{26,27,28} to atoms \cite{29,30,31} and ions \cite{32,33}, and from nitrogen-vacancy
centers \cite{34,35,36} to superconducting circuits \cite{37,38,39}. However, these
experiments have been confined to realizations of NH semiclassical models,
where the degree of freedom either of the light or of the matter was treated
classically in the effective NH Hamiltonian describing the light-matter
interaction, and consequently, the observed NH effects bear no relation to
quantum entanglement. Indeed, all the genuinely NH effects demonstrated so far
can occur in both quantum and classical systems. This naturally leads to an
imperative issue: What feature can simultaneously manifest non-Hermiticity
and non-classicality?
\textcolor{black}{Recently, there has been a significant focus on nonequilibrium quantum phase transitions in the entanglement dynamics of NH many-body systems \cite{40,41,42}, but the purely quantum-mechanical NH features associated with the Hamiltonian eigenstates have remained unexplored.}

\textcolor{black}{We here perform an in-depth investigation on this fundamental problem, finding that dissipative interacting quantum systems can display exceptional entanglement phenomena. In particular, we discover an EP-induced entanglement transition, which represents a purely quantum-mechanical NH signature, with neither Hermitian nor classical analogs. We illustrate our discovery with a dissipative qubit-photon system, whose NH quantum effects are manifested by the singular entanglement behaviors of the bipartite entangled eigenstates. We experimentally demonstrated these singular behaviors in a circuit, where a superconducting qubit is controllably coupled to a decaying resonator. The exceptional entanglement signatures, inherent in the qubit-photon static eigenstates, are mapped out by a density matrix post-casting method, which enables us to extract the weak nonclassical signal from the strong noise background. In addition to the quantum character, the demonstrated NH phenomena originate from naturally occurring dissipation, distinct from that induced by an artificially engineered reservoir \cite{12,13,14,15,16,17,18,19,20,21,22,23,24,25,26,27,28,29,30,31,32,33,34,35,36,37,38,39}. Our results are universal for composite quantum systems immersed in pervasive Markovian reservoirs, endowing NH quantum mechanics with genuinely non-classical characters, which are absent in NH classical physics.}

The theoretical model, used to illustrate the NH entanglement transition, is
composed of a two-level system (qubit) resonantly coupled to a quantized
photonic mode, as sketched in Fig. 1a. The quantum state evolution
trajectory without photon-number jumps is governed by the NH Hamiltonian
(setting $\hbar =1$) 
\begin{equation}
{\cal H}_{NH}=\Omega (a^{\dagger }\left\vert g\right\rangle \left\langle
e\right\vert +a\left\vert e\right\rangle \left\langle g\right\vert )-\frac{i%
}{2} \kappa _{q}\left\vert e\right\rangle \left\langle e\right\vert -%
\frac{i}{2} \kappa _{f}a^{\dagger }a,
\end{equation}%
where $\left\vert e\right\rangle $ $(\left\vert g\right\rangle) $ denotes the upper (lower) level of the qubit, $a^{\dagger }$
$(a)$ represents the creation (annihilation) operator for the photonic
mode, $\kappa _{q}$ $(\kappa _{f})$ is the energy dissipation
rate for the qubit (field mode), and $\Omega $ is the qubit-field
coupling strength. In the $n$-excitation subspace, the system has two \textcolor{black}{right entangled eigenstates},
given by

\begin{equation}
\left\vert \Phi_{n,\pm }\right\rangle ={\cal N}_{n,\pm }(\sqrt{n}\Omega \left\vert
e,n-1\right\rangle +E_{n,\pm}\left\vert g,n\right\rangle ),
\end{equation}%
where \textcolor{black}{${\cal N}_{n,\pm }=(n\Omega^2+|E_{n,\pm}|^2)^{-1/2}$} and \textcolor{black}{$E _{n,\pm }=-i\gamma /4\pm \Delta E_{n}/2$ are the corresponding eigenenergies}, with $\gamma =\kappa _{f}+\kappa _{q}$, \textcolor{black}{$\Delta E_{n}=2\sqrt{n\Omega ^{2}-\kappa ^{2}/16}
$}, and $\kappa
=\kappa _{f}-\kappa _{q}$. These two eigenstates
are separated by an energy gap of $\Delta E_{n}$. The inherent quantum
entanglement makes the system fundamentally distinct from previously
demonstrated NH semiclassical models \cite{29,30,31,32,33,34,35,36,37,38,39}, where the qubit is not entangled with the classical control field in any way. When $\Omega >\kappa
/(4\sqrt{n})$, the system has a real energy gap, and undergoes Rabi-like
oscillations, during which the qubit periodically exchanges a photon with the
field mode. With the decrease of $\Omega $, the energy gap is continuously
narrowed until reaching the EP, where the two energy levels coalesce.
After crossing the EP, the gap becomes imaginary, and the population
evolution exhibits an over-damping feature.

Unlike previous investigations, here each eigenenergy is possessed by the
two entangled components, neither of which has its own state. 
%The nonclassical feature of the composite system is manifested by the quantum
%entanglement of the underlying eigenstate, 
\textcolor{black}{The nonclassical feature of each eigenstate is manifested by the light-matter entanglement,}
which can be quantified by the concurrence \cite{43} 
\begin{equation}
{\cal E}_{\pm }=\frac{2\sqrt{n}\Omega \left\vert E_{n,\pm}\right\vert }{%
\left\vert E_{n,\pm}\right\vert ^{2}+n\Omega ^{2}}.
\end{equation}%
When the rescaled coupling strength $\eta =4\sqrt{n}\Omega /\kappa \ $is
much smaller than 1, the two eigenstates are respectively dominated by $%
\left\vert e,n-1\right\rangle $ and $\left\vert g,n\right\rangle $. With the
increase of $\eta $, these two populations become increasingly balanced
until reaching the EP $\eta =1$, where both eigenstates converge to the same
maximally entangled state. During this convergence, ${\cal E}_{\pm }$
exhibit a linear scaling with $\eta $. In the Bloch representation, this
corresponds to a rotation of the eigenvector $\left\vert \Phi_{n,+}\right\rangle $ ($\left\vert \Phi_{n,-}\right\rangle $) around the $x$ ($-x$)
axis from pointing at the north (south) polar, until merging at the $y$
axis, as shown in the left panel of Fig. 1b. After crossing the EP, ${\cal E}%
_{\pm }$ become independent of $\Omega $, which implies both two eigenstates
keep maximally entangled. However, $\left\vert \Phi_{n,+}\right\rangle $ ($%
\left\vert \Phi_{n,-}\right\rangle $) is rotated around the $z$ ($-z$) axis,
progressively approaching the $x$ ($-x$) axis (right panel of Fig. 1b). This
sudden switch of the rotation axis manifests an entanglement transition at
the EP, where the derivative of the concurrence with respect to $\eta $
presents a discontinuity, jumping from $1$ to $0$.
By measuring the time-evolving output states associated with the
no-jump trajectory, we can extract the information about both the energy gap
and entanglement regarding the ``static'' eigenstates. It should be noted
that the resonator plays a radically different role from the ancilla used in
the previous experiments \cite{34,35,36,37}, which was introduced as an artificially
engineered environment to the test qubit, but whose dynamics was not
included in the effective NH Hamiltonian dynamics. In distinct
contrast, in the present system the degree of freedom of $R$ constitutes a
part of the Hamiltonian, whose NH term is induced by the natural environment.

The experimental demonstration of the NH physics is
performed in a circuit quantum electrodynamics architecture, where the
qubit-photon model is realized with a superconducting qubit $Q$ and its
resonator $R$ with a fixed frequency $\omega _{r}/2\pi=6.66$ GHz (see Supplemental Material, Sec. S2 \cite{45}). The decaying rates of $%
Q$ and $R$ are $\kappa _{q}\simeq 0.07$ MHz and $\kappa _{f}= 5$ MHz, respectively. The
accessible maximum frequency of $Q$, $\omega _{\max }=2\pi\times6.01$ GHz, is lower than $%
\omega _{r}$ by an amount much larger than their on-resonance swapping
coupling $g_r=2\pi\times41$ MHz (Fig. 1c). To observe the EP physics, an ac flux is
applied to $Q$, modulating its frequency as $\omega _{q}=\omega
_{0}+\varepsilon \cos (\nu t)$, where $\omega _{0}$ is the mean $\left\vert
e\right\rangle $-$\left\vert g\right\rangle $ energy difference, and $%
\varepsilon $ and $\nu $ denote the modulating amplitude and frequency. This
modulation enables $Q$ to interact with $R$ at a preset sideband, with the
photon swapping rate $\Omega $ tunable by $\varepsilon $ (see Supplemental Material, Sec. S3 \cite{45}).

Our experiment focuses on the single-excitation case ($n=1$). Before the experiment, the system is initialized to the ground state $\left\vert
g,0\right\rangle $. The experiment starts by transforming $Q$ from the
ground state $\left\vert g\right\rangle $\ to excited state $\left\vert
e\right\rangle $\ with a $\pi $ pulse, following which the parametric
modulation is applied to $Q$\ to initiate the $Q$-$R$\ interaction \textcolor{black}{(see Fig. S4 of Supplementary Material for the pulse sequence)}. This interaction, together with the natural dissipations, realizes the
NH Hamiltonian of Eq. (1). After the modulating pulse, the $Q$-$R$\ state is
measured with the assistance of an ancilla qubit ($Q_{a}$) and a bus
resonator ($R_{b}$), which is coupled to both $Q$\ and $Q_{a}$. The
subsequent $Q\rightarrow R_b$, $R_b\rightarrow Q_{a}$, and $R\rightarrow Q$\
quantum state transferring operations map the $Q$-$R$\ output state to the $%
Q_{a}$-$Q$\ system, whose state can be read out by quantum state tomography
(see Supplemental Material, Sec. S5 \cite{45}).

%The population evolutions coinciding with the NH Hamiltonian dynamics is
%post-selected by discarding the joint $Q_{a}$-$Q$ detection events $%
%\left\vert g,g\right\rangle $. 
\textcolor{black}{A defining feature of our system is the conservation of the excitation number under the NH Hamiltonian. This Hamiltonian evolves the system within the subspace $\{|e,0\rangle,|g,1\rangle\}$. A quantum jump would disrupt this conversion, moving the system out of this subspace. This property in turn enables us to post-select the output state governed the NH Hamiltonian simply by discarding the joint $Q_a$-$Q$ outcome $|g,g\rangle$ after the state mapping.}
With a correction of the quantum state
distortion caused by the decoherence occurring during the state mapping, we
can infer the quantum Rabi oscillatory signal, characterized by the
evolution of the joint probability $\left\vert e,0\right\rangle $, denoted
as $P_{e,0}$. Fig. 2a shows thus-obtained $P_{e,0}$ as a function of the
rescaled coupling $\eta $ and the $Q$-$R$ interaction time $t$. The results
clearly show that both the shapes and periods of vacuum Rabi oscillations
are significantly modulated by the NH term around the EP $\eta =1$, where
incoherent dissipation is comparable to the coherent interaction. These
experimental results are in well agreement with numerical simulations (see Supplemental Material, Sec. S4 \cite{45}).

The Rabi signal does not unambiguously reveal the system's quantum behavior.
To extract full information of the two-qubit entangled state, it is
necessary to individually measure all three Bloch vectors for each
qubit, and then correlate the results for the two qubits. The $z$-component can be directly
measured by state readout, while measurements of the $x$- and $y$-components
require $y$-rotations and $x$-rotations before state readout, which breaks down
excitation-number conservation, and renders it
impossible to distinguish individual jump events from no-jump ones. We
circumvent this problem by first reconstructing the two-qubit density matrix
using all the measurement outcomes, and then discarding the matrix elements
associated with $\left\vert g,g\right\rangle $ (see Supplemental Material, Sec. S5 \cite{45}). This effectively post-casts
the two-qubit state to the subspace \{$\left\vert e,g\right\rangle $, $%
\left\vert g,e\right\rangle $\}. We note that such a technique is in
sharp contrast with the conventional post-selection method, where some
auxiliary degree of freedom (e.g., propagation direction of a photon) enables
reconstruction of the relevant conditional output state, but which is
unavailable in the NH qubit-photon system. With a proper correction for the
state mapping error, we obtain the $Q$-$R$ output state governed by the
non-Hermitian Hamiltonian. In Fig. 2b, we present the resulting $Q$-$R$
concurrence, as a function of $\eta $ and $t$. To show the entanglement
behaviors more clearly, we present the concurrence evolutions for $\eta =5$
and $0.5$ in Fig. 2c and d, respectively. The results demonstrate that the
entanglement exhibits distinct evolution patterns in the regimes above and
below the EP.

To reveal the close relation between the exceptional entanglement
behavior and the EP, we infer the eigenvalues and eigenstates of the NH
Hamiltonian from the output states, measured for different evolution times (see Supplemental Material, Sec. S7 \cite{45}).
The concurrences associated with the two eigenstates, obtained for different
values of $\eta $, are shown in Fig. 3a and b (dots), respectively. As
expected, each of these entanglements exhibits a linear scaling below the
EP, but is saturated at the EP, and no longer depends upon $\eta $ \textcolor{black}{after} crossing the EP. The singular features around the EP are manifested by their
derivatives to $\eta $, which are shown in the insets. The discontinuity of
these derivatives indicates the occurrence of an entanglement transition at the
EP. Such a nonclassical behavior, as a consequence of the competition
between the coherent coupling and incoherent dissipation, represents a
unique character of strongly correlated NH quantum systems, but has not been
reported so far. These results demonstrate a longitudinal merging of the two
entangled eigenstates at the EP (left panel of Fig. 2b). Accompanying this
is the onset of the transverse splitting, which can be characterized by
the relative phase difference between $\left\vert \Phi_{1,\pm}\right\rangle $%
, defined as $\varphi =$ $\varphi _{+}-\varphi _{-}$, with $\varphi _{\pm }$
representing the relative phases between $\left\vert g,1\right\rangle $ and $%
\left\vert e,0\right\rangle $ in the eigenstates $\left\vert \Phi_{1,\pm
}\right\rangle $. Such a phase difference, inferred for different values of $%
\eta $, is displayed in Fig. 3c.
\textcolor{black}{This quantum-mechanical NH signature is universally applicable to dissipative interacting quantum systems, a feature absent in earlier superconducting-circuit-based single-qubit NH experiments \cite{38,39}.}
%In a recent two-qubit experiment, entanglement generation was  accelerated by non-Hermiticity \cite{47}; however, the aspect of entanglement transition remained unexplored.

The demonstrated exceptional entanglement transition is a
purely quantum-mechanical NH effect. On one hand, quantum entanglement,
which has no classical analogs, represents \textcolor{black}{a most} characteristic trait
that distinguishes quantum physics from classical mechanics \cite{48}. On the other hand, this effect, uniquely associated with the EP,
is absent in Hermitian qubit-photon systems \cite{49,50,51,52}. We further note that 
such an NH effect can occur in other dissipative quantum-mechanically
correlated systems, e.g., a system composed of two or more coupled qubits
with unbalanced decaying rates (see Supplemental Material, Sec. S8 \cite{45}). This
implies that such a quantum-mechanical NH signature is universal for
dissipative interacting quantum systems.

The real and imaginary parts of the extracted quantum Rabi splitting \textcolor{black}{$\Delta E_{n}$}
are shown in Fig. 3d. As theoretically predicted, above the EP the two
eigenenergies have a real gap, which is continuously narrowed when the
control parameter $\eta $ is decreased until reaching the EP, where the two
levels coalesce. After crossing the EP, the two levels are re-split but with
an imaginary gap, which increases when $\eta $ is decreased. This
corresponds to the real-to-imaginary transition of the vacuum Rabi splitting
between the $Q$-$R$ entangled eigenstates, which is in distinct contrast
with the previous experiments on ${\cal PT}$ symmetry breaking \cite{29,30,31,32,33,34,35,36,37,38,39},
realized with the semiclassical models, where the exceptional physics has no
relation with quantum entanglement.

In conclusion, we have discovered an exceptional entanglement transition in
a fundamental light-matter system governed by an NH Hamiltonian, establishing
a close connection between quantum correlations and non-Hermitian effects.
This transition has neither Hermitian nor classical analogs, representing
the unique feature of NH quantum mechanics. The experimental demonstration
is performed in a circuit, where a superconducting qubit is controllably
coupled to a resonator with a non-negligible dissipation induced by a
natural Markovian reservoir. The NH quantum signatures of the eigenstates
are inferred from the no-jump output state, measured for different evolution
times. Our results push NH Hamiltonian physics from the classical to
genuinely non-classical regime, where the emergent phenomena have neither
Hermitian nor classical analogs. 
%The post-projection method, developed for
%extracting the no-jump output density matrix, would open the door to
%experimentally explore purely quantum-mechanical NH effects in interacting
%systems.
%\textcolor{black}{The post-projection method, developed for extracting the no-jump output density matrix, would open the door to experimentally explore purely quantum-mechanical NH effects in a broad spectrum of interacting systems, where excitation number is initially definite and conserved under the NH Hamiltonian. Such systems include resonator-qubits arrays \cite{53}, fully-connected architectures \cite{54} and lattice models \cite{55}. When the system initially has $n$ excitations, the no-jump trajectory can be post-selected by discarding the outcomes with less than $n$ excitations.}
\textcolor{black}{The post-projection method, developed for extracting the no-jump output density matrix, would open the door to experimentally explore purely quantum-mechanical NH effects in a broad spectrum of interacting systems, where excitation number is initially definite and conserved under the NH Hamiltonian. Such systems include resonator-qubits arrays \cite{53}, fully-connected architectures involving multiple qubits coupled to a single resonator \cite{54}, and lattices composed of many qubits with nearest-neighbor coupling \cite{55}. When the system initially has $n$ excitations, the no-jump trajectory can be post-selected by discarding the outcomes with less than $n$ excitations.}

\begin{acknowledgments}
Acknowledgments: We thank Liang Jiang at University of Chicago for valuable
comments. Funding: This work was supported by the National Natural Science
Foundation of China (Grant No. 12274080, No. 11875108, No. 12204105, No.
11774058, No. 12174058, No. 11974331, No. 11934018, No.
92065114, and No. T2121001), Innovation Program for Quantum Science and
Technology (Grant No. 2021ZD0300200 and No. 2021ZD0301200), NSF (Grant No. 2329027), the Strategic Priority Research Program
of Chinese Academy of Sciences (Grant No. XDB28000000), the National Key
R\&D Program under (Grant No. 2017YFA0304100), the Key-Area Research and
Development Program of Guangdong Province, China (Gran No. 2020B0303030001),
Beijing Natural Science Foundation (Grant No. Z200009), and the Project from
Fuzhou University (Grant No. 049050011050). 

Author contributions: S.B.Z. predicted the exceptional entanglement
transition and conceived the experiment. P.R.H. and X.J.H., supervised by
Z.B.Y. and S.B.Z., carried out the experiment. F.W., P.R.H., J.W., and
S.B.Z. analyzed the data. S.B.Z., J.W., Z.B.Y., and W.Y. cowrote the paper.
All authors contributed to interpretation of observed phenomena and helped
to improve presentation of the paper.
\end{acknowledgments}

\begin{figure*}[htbp]
	\centering
	\includegraphics[width=7in]{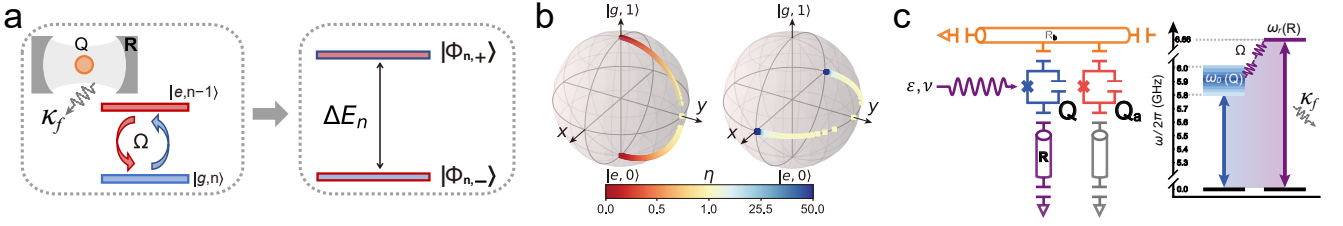}
	\caption{\textbf{NH Light-matter interaction and experimental implementation.} (a) Theoretical model. The system under investigation
		involves a quantized field mode with a decaying rate $\kappa_f $ resonantly
		interacting with a matter qubit with a decaying rate $\kappa_q$. Their
		interaction is characterized by a coupling strength of $\Omega $. The
		system's dynamics can be cast onto different U(1)-symmetric eigenspaces,
		denoted as $\{\left\vert e,n-1\right\rangle ,\left\vert g,n\right\rangle \}$%
		, within each of which there exist two eigenstates $\left\vert \Phi_{n,\pm
		}\right\rangle $, separated by a gap of \textcolor{black}{$\Delta E_{n}$}. (b) Bloch representation of 
		$\left\vert \Phi_{1,\pm}\right\rangle $ in the single-excitation subspace ($%
		n=1$). For clarity, we here pictorially display the dependence of $%
		\left\vert \Phi_{1,\pm}\right\rangle $ on the rescaled coupling strength $%
		\eta =4\Omega /\kappa$. For $\eta =0$, $\left\vert \Phi_{1,\pm
		}\right\rangle $ correspond to $\left\vert e,0\right\rangle $ and $%
		\left\vert g,1\right\rangle $, respectively. With the increase of $\eta $, $%
		\left\vert \Phi_{1,\pm}\right\rangle $ are respectively rotated around $\pm
		x $ axes, and merged to the $y$ axis at the EP $\eta =1$, featuring the occurrence of the maximally entangled state \textcolor{black}{$\left\vert Y\right\rangle
			=(\left\vert g,1\right\rangle -i\left\vert e,0\right\rangle )/\sqrt{2}$}.
		After crossing the EP, the eigenvectors $\left\vert \Phi_{1,\pm
		}\right\rangle $ remain on the equatorial plane, but are rotated around the $%
		z$ and $-z$ axes, progressively aligned with the $\pm x$ axes, tending
		to \textcolor{black}{$\left\vert \pm X\right\rangle =(\left\vert g,1\right\rangle \pm
			\left\vert e,0\right\rangle )/\sqrt{2}$}, respectively. (c) Implementation of
		the NH Hamiltonian. In the experimental system, \textcolor{black}{a superconducting} qubit \textcolor{black}{$Q$ is} highly
		detuned from the lossy resonator $R$ with a fixed frequency $\omega _{r}/2\pi=6.66$ GHz.
		The $Q$-$R$ interaction is enabled with an ac flux, which modulates $Q$'s
		energy gap around the mean value $\omega _{0}$, with a frequency $\nu $,
		mediating a photonic swapping coupling at one sideband, with the coupling
		strength controlled by the modulating amplitude $\varepsilon $. 
%		(d) Pulse sequence. The NH dynamics starts with the initial state $\left\vert e,0\right\rangle $, prepared from $\left\vert g,0\right\rangle $ with a $\pi$ pulse. After a preset evolution time, the parametric modulation is switched off, followed by the state mappings $Q\rightarrow R_{b}$, $R_b\rightarrow Q_a$, and $R\rightarrow Q$, each realized by an on-resonance swapping gate, where $Q_{a}$ is an ancilla qubit, and $R_{b}$
%		represents the bus resonator coupled to both qubits. After these mappings, the $Q$-$R$ output state produced by the NH dynamics is encoded in the joint $Q_{a}$-$Q$ state, which is then measured by quantum state tomography.
	}
	\label{Fig1}
\end{figure*}

\begin{figure*}[htbp]
	\centering
	\includegraphics[width=6in]{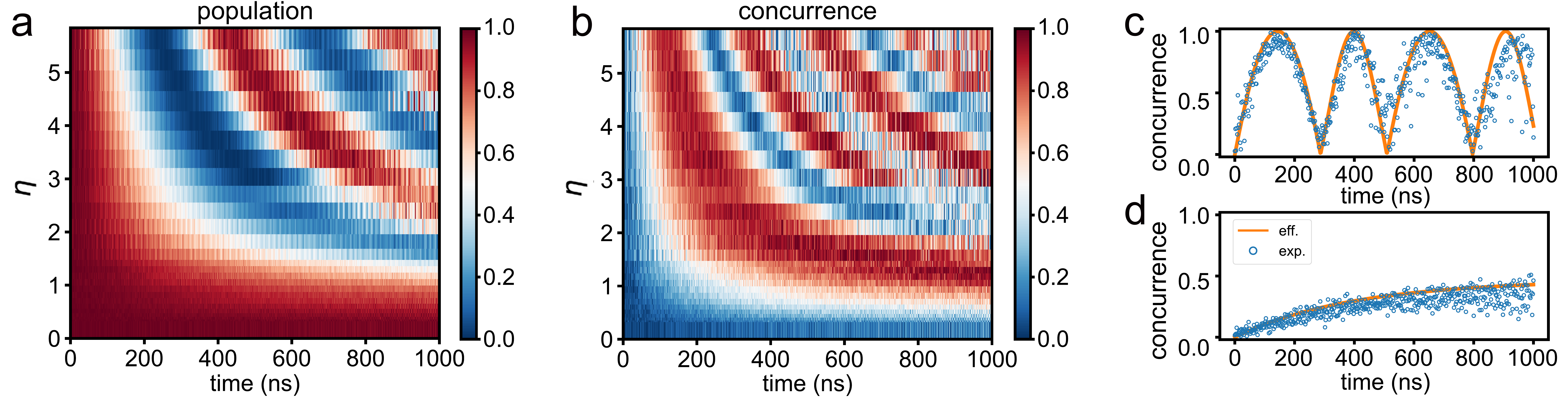}
	\caption{\textbf{Characterization of NH dynamical evolutions.} (a)
		\textcolor{black}{Measured evolutions of the population $|e,0\rangle$} for different values of the rescaled coupling $\eta 
		$. The signals are extracted by correlating the measurement outcomes of $Q$
		and $Q_{a}$, \textcolor{black}{which correspond to the outputs of $R$ and $Q$ right before the state mappings.} The NH Hamiltonian evolution trajectory is post-selected by
		discarding the detection events $\left\vert g,g\right\rangle $, and
		renormalizing the remaining outcomes $\left\vert g,e\right\rangle $ and $%
		\left\vert e,g\right\rangle $. (b) Evolutions of the concurrences for
		different values of $\eta $. The concurrence at each point is inferred with
		the post-projected density matrix, obtained by reconstructing the output $Q$-%
		$Q_{a}$ density matrix, projecting \textcolor{black}{the} thus-obtained two-qubit state to the
		reduced subspace with one excitation, and then renormalizing the remaining
		matrix elements. (c), (d) Concurrence evolutions for $\eta =5$ (c) and $0.5$
		(d). The orange solid curves are the numerical simulations
		using the effective Hamiltonian and the blue circles are the experimental results.}
	\label{Fig2}
\end{figure*}

\begin{figure*}[htbp]
	\centering
	\includegraphics[width=7in]{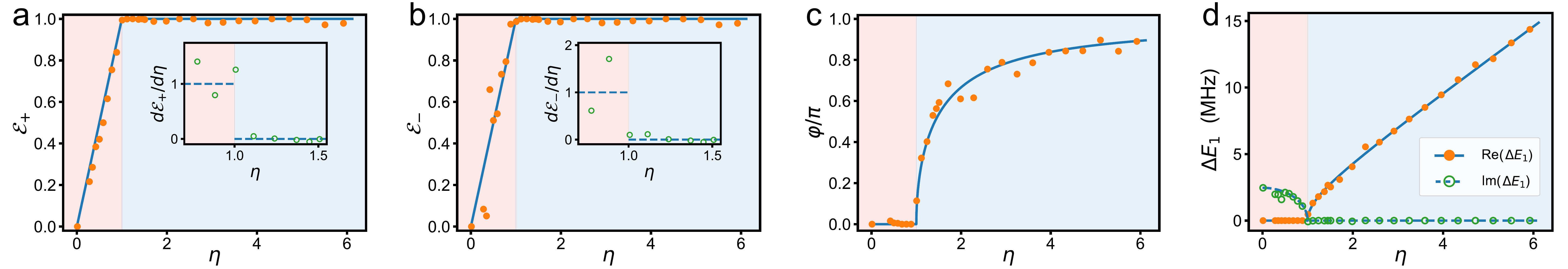}
	\caption{\textbf{Observation of exceptional phase transitions.} (a),
		(b) \textcolor{black}{Entanglements} for the eigenstates $\left\vert \Phi_{1,+}\right\rangle $ and $\left\vert \Phi_{1,-}\right\rangle $ versus $\eta $.
		We obtain the concurrences ${\cal E}_{\pm }$ at each point based on the
		corresponding eigenstates, which are mapped out from the \textcolor{black}{$Q$-$R$} output
		states, measured for different evolution times. The dots denote the inferred
		concurrences, \textcolor{black}{which well agree with the results for the ideal output states}. The derivatives $d{\cal E}%
		_{\pm }/d\eta $ around the EP, \textcolor{black}{obtained by} $[{\cal E}_{\pm }(\eta
		+\delta \eta )-{\cal E}_{\pm }(\eta )]/\delta \eta $, is displayed in the
		insets. (c) Relative phase difference ($\varphi $)
		between $\left\vert \Phi_{1,\pm}\right\rangle $. This phase difference is
		defined as $\varphi =\varphi _{+}-\varphi _{-}$, where $\varphi _{\pm }$ are
		the relative phase between $\left\vert g,1\right\rangle $ and $\left\vert
		e,0\right\rangle $ in the eigenstates $\left\vert \Phi_{1,\pm}\right\rangle $%
		. (d) Spectral gap \textcolor{black}{$\Delta E_{1}$}. As the eigenspectrum is possessed by the
		entangled eigenstates of the $Q$-$R$ system, this gap corresponds to the
		vacuum Rabi splitting. The solid and dashed lines denote the real and
		imaginary parts, respectively.}
	\label{Fig3}
\end{figure*}

\clearpage

%\begin{references}
\bibliographystyle{unsrt}

\end{document}

\title{Supplementary Materials for \protect\\``Exceptional entanglement phenomena: non-Hermiticity meeting
	non-classicality"}
\author{Pei-Rong Han$^{1}$}
\thanks{These authors contribute equally to this work.}
\author{Fan Wu$^{1}$}
\thanks{These authors contribute equally to this work.}
\author{Xin-Jie Huang$^{1}$}
\thanks{These authors contribute equally to this work.}
\author{Huai-Zhi Wu$^{1}$}
\author{Chang-Ling Zou$^{2,3,8}$, Wei Yi$^{2,3,8}$, Mengzhen Zhang$^{4}$, Hekang Li$^{5}$%
	, Kai Xu$^{5,6,8}$, Dongning Zheng$^{5,6,8}$, Heng Fan$^{5,6,8}$}
\author{Jianming Wen$^{7}$}
\thanks{E-mail: jianming.wen@kennesaw.edu}
\author{Zhen-Biao Yang$^{1,8}$}
\thanks{E-mail: zbyang@fzu.edu.cn}
\author{Shi-Biao Zheng$^{1,8}$}
\thanks{E-mail: t96034@fzu.edu.cn}
\address{$^{1}$Fujian Key Laboratory of Quantum Information and Quantum Optics, College of Physics and Information Engineering, Fuzhou University, Fuzhou, Fujian, 350108, China\\
	$^{2}$CAS Key Laboratory of Quantum Information, University of Science and\\
	Technology of China, Hefei 230026, China\\
	$^{3}$CAS Center for Excellence in Quantum Information and Quantum Physics,\\
	University of Science and Technology of China, Hefei 230026, China\\
	$^{4}$Pritzker School of Molecular Engineering, University of Chicago,\\
	Chicago, IL 60637, USA\\
	$^{5}$Institute of Physics, Chinese Academy of Sciences, Beijing 100190,\\
	China\\
	$^{6}$CAS Center for Excellence in Topological Quantum Computation,\\
	University of Chinese Academy of Sciences, Beijing 100190, China\\
	$^{7}$Department of Physics, Kennesaw State University, Marietta, Georgia\\
	30060, USA\\
	$^{8}$Hefei National Laboratory, Hefei 230088, China}
\maketitle

\tableofcontents
\section{Entanglement of the Eigenstates and NH-Hamiltonian-Evolved State}
When the qubit-resonator system is restricted in the $n$-excitation subspace,
the resonator can be thought of as a qubit with two basis vectors $\left\vert
n-1\right\rangle $ and $\left\vert n\right\rangle $. With this analogy, the
qubit-resonator model corresponds to a composite system composed of two
qubits, whose entanglement can be quantified in terms of the eigenvalues of
the operator \cite{5}
\begin{equation}
\overset{\sim }{\rho }=\rho (\sigma _{y}^{q}\otimes \sigma _{y}^{r})\rho
^{\ast }(\sigma _{y}^{q}\otimes \sigma _{y}^{r}),
\end{equation}%
where $\rho $ is the density operator of the composite system, and $\sigma
_{y}^{q}$ and $\sigma _{y}^{r}$ denote the corresponding $y$-component Pauli
operators of the two qubits, defined as 
\begin{eqnarray}
	\sigma _{y}^{q} &=&-i\left\vert g\right\rangle \left\langle e\right\vert
	+i\left\vert e\right\rangle \left\langle g\right\vert , \\
	\sigma _{y}^{r} &=&-i\left\vert n-1\right\rangle \left\langle n\right\vert
	+i\left\vert n\right\rangle \left\langle n-1\right\vert .
\end{eqnarray}%
Suppose that $\lambda _{1}\geq \lambda _{2}$ $\geq \lambda _{3}\geq \lambda
_{4}$ are the square roots of the eigenvalues of $\overset{\sim }{\rho }$.
Then the two-qubit entanglement associated with the density matrix $\rho $
is measured by the quantity 
\begin{equation}
\mathcal{E}=\max \{\lambda _{1}-\lambda _{2}-\lambda _{3}-\lambda _{4}\text{%
	, }0\}.\text{ }
\end{equation}
$\mathcal{E}$ is referred to as concurrence \cite{5}, ranging from $0$ to $1$.

For the eigenstates $\left\vert \Phi _{n,\pm }\right\rangle $ of the NH
Hamiltonian, given by Eq. (2) of the main text, the system density operator
in the basis $\left\{ \left\vert g,n-1\right\rangle ,\left\vert
g,n\right\rangle ,\left\vert e,n-1\right\rangle ,\left\vert e,n\right\rangle
\right\} $ can be expressed as 
\begin{equation}
\rho _{n,\pm }=\left\vert \mathcal{N}_{n,\pm }\right\vert ^{2}\left( 
\begin{array}{cccc}
0 & 0 & 0 & 0 \\ 
0 & \left\vert E_{n,\pm }\right\vert ^{2} & \sqrt{n}\Omega E_{n,\pm } & 0 \\ 
0 & \sqrt{n}\Omega E_{n,\pm }^{\ast } & n\Omega ^{2} & 0 \\ 
0 & 0 & 0 & 0%
\end{array}%
\right) .
\end{equation}%
The corresponding matrix $\overset{\sim }{\rho }$ has a single non-zero
eigenvalue, given by $4n\left\vert \Omega E_{n,\pm }\right\vert ^{2}\mathcal{%
	N}_{n,\pm }^{4}$. The resulting concurrence is%
\begin{eqnarray}
	\mathcal{E}_{\pm } &=&2\sqrt{n}\Omega \left\vert E_{n,\pm }\right\vert 
	\mathcal{N}_{n,\pm }^{2} \\
	&=&\frac{2\sqrt{n}\Omega \left\vert E_{n,\pm}\right\vert }{%
		\left\vert E_{n,\pm}\right\vert ^{2}+n\Omega ^{2}}.
\end{eqnarray}

\textcolor{blue}{When $\eta =4\Omega /\kappa <1$, the concurrence of each eigenstate increases linearly with $\eta $ until reaching the maximum 1 at the EP. After crossing the EP, the concurrence becomes independent of $\eta $. This exceptional entanglement transition can be elucidated as follows. The entanglement between the qubit and the photonic mode arises from the coherent superposition of the two basis vectors $\left\vert e,n-1\right\rangle $ and $\left\vert g,n\right\rangle $. The amount of entanglement hinges on the relative weighting of these basis vectors. Below the EP, their populations are unequal in each eigenstate. As $\eta$ increases, these populations gradually balance until $\eta=1$. Beyond this point, further increments in $\eta$ only change the relative phase of the two superimposed basis vectors, while their populations remain evenly distributed.} The entanglement can also be characterized by the negativity of the partial
transpose of the density matrix \cite{6}. For each partial
transpose, the negativity is defined as the absolute value of the sum of the
negative eigenvalues. The partial transposes corresponding to the two
eigenstates are 
\begin{equation}
\rho _{n,\pm }^{T}=\mathcal{N}_{n,\pm }^{2}\left( 
\begin{array}{cccc}
0 & 0 & 0 & \sqrt{n}\Omega E_{n,\pm }^{\ast } \\ 
0 & \left\vert E_{n,\pm }\right\vert ^{2} & 0 & 0 \\ 
0 & 0 & n\Omega ^{2} & 0 \\ 
\sqrt{n}\Omega E_{n,\pm } & 0 & 0 & 0%
\end{array}%
\right) .
\end{equation}%
The corresponding negativities are $\sqrt{n}\Omega \left\vert E_{n,\pm
}\right\vert \mathcal{N}_{n,\pm }^{2}$, each of which is equal to half of the
corresponding concurrence. This implies that there is a monotonous
one-to-one correspondence between concurrence and negativity, which ranges
from 0 to 1/2.

In our experiment, the eigenstates are extracted from the output state
associated with the no-jump trajectory, measured for different interaction
times. The system starts from the initial state $\left\vert e,0\right\rangle 
$. After an interaction time $t$, the system state, evolved under the NH
Hamiltonian, can be expressed as a linear combination of the two eigenstates,%
\begin{equation}
\left\vert \psi _{n}(t)\right\rangle =K_{\pm }(t)\left( \frac{e^{-iE_{n,+}t}%
}{\mathcal{N}_{n,+}E_{n,+}}\left\vert \Phi _{n,+}\right\rangle -\frac{%
	e^{-iE_{n,-}t}}{\mathcal{N}_{n,-}E_{n,-}}\left\vert \Phi _{n,-}\right\rangle
\right) ,
\end{equation}%
where 
\begin{equation}
K_{\pm }(t)=\left( \left\vert \frac{e^{-iE_{n,+}t}}{\mathcal{N}_{n,+}E_{n,+}}%
\right\vert ^{2}+\left\vert \frac{e^{-iE_{n,-}t}}{\mathcal{N}_{n,-}E_{n,-}}%
\right\vert ^{2}\right) ^{-1/2}.
\end{equation}%
Replacing Eq. (2) of the main text into this linear combination of $%
\left\vert \Phi _{n,\pm }\right\rangle $, we can obtain the state evolution
in terms of the basis vectors $\left\vert g,n\right\rangle $ and $\left\vert
e,n-1\right\rangle $, given by
\begin{eqnarray}
\left\vert \psi _{n}(t)\right\rangle &=&{\cal N}_{n}\{[2\Delta E_{n}\cos
(\Delta E_{n}t/2)+\kappa \sin (\Delta E_{n}t/2)]\left\vert e,n-1\right\rangle \\
&&-i4\sqrt{n}\Omega \sin (\Delta E_{n}t/2)\left\vert g,n\right\rangle \},  \nonumber
\end{eqnarray}
%where ${\cal N}_n = e^{-\kappa t/4}/(2\Delta E_n)$.
where ${\cal N}_n = (|2\Delta E_{n}\cos
(\Delta E_{n}t/2)+\kappa \sin (\Delta E_{n}t/2)|^2+|4\sqrt{n}\Omega \sin (\Delta E_{n}t/2)|^2)^{-1/2}$ is the normalization factor.
For the state $%
\left\vert \psi _{n}(t)\right\rangle $, the qubit-resonator concurrence is
given by%
\begin{equation}
\mathcal{E}=\sin (2\theta ),
\end{equation}%
where%
\begin{equation}
\theta =\arctan \left\vert \frac{4\sqrt{n}\Omega \sin (\Delta E_{n}t/2)}{%
	2\Delta E_{n}\cos (\Delta E_{n}t/2)+\kappa \sin (\Delta E_{n}t/2)}%
\right\vert .
\end{equation}

\section{Experimental setup and system parameters}
\begin{figure*}[htbp] 
	\centering
	\includegraphics[width=7in]{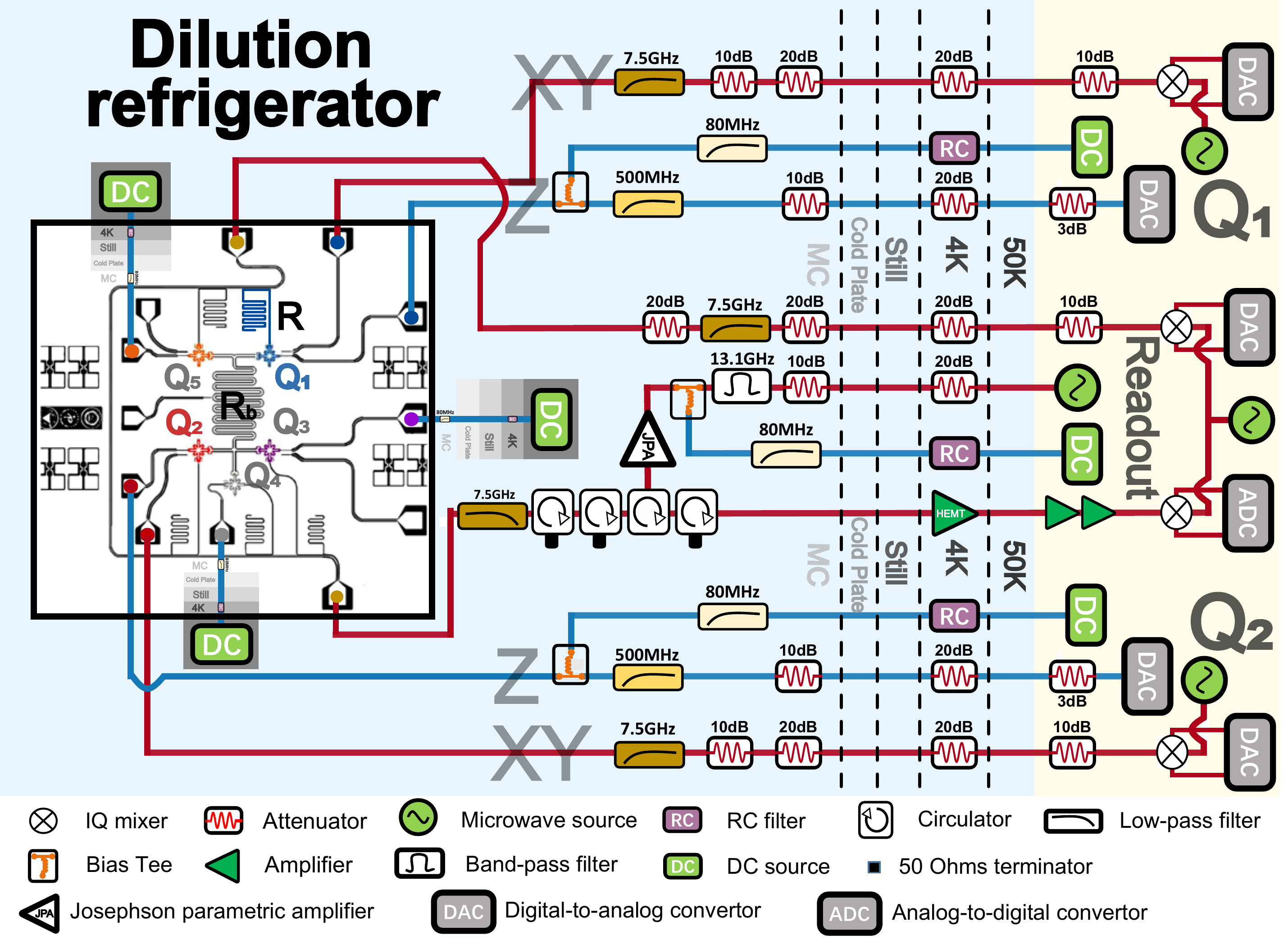}
	\caption{\textbf{Schematic layouts of our circuit QED system and experimental setup.} \textcolor{black}{The superconducting circuit has five frequency-tunable Xmon qubits, labeled from $Q_1$ to $Q_5$.} Each qubit can be individually frequency-biased and frequency-modulated (through the Z line) and flipped (through the XY line). Thanks to such flexible adjustability, every qubit \textcolor{black}{can} be coupled to the bus resonator ($R_b$) in a controllable way. The XY control of each qubit is implemented by mixing the low-frequency signals (yielded by two Digital-to-analog converter (DAC)'s I/Q channels) with a Microwave source (MS) at 5.5-GHz carrier frequency; while the Z control is fulfilled by two signals: one is produced by the Direct-current (DC) biasing line from a low frequency DC source, and the other is directly \textcolor{black}{obtained from} the Z control of a DAC. Meantime, every qubit has its own readout resonator which helps to project out the state information. Experimentally, this is accomplished by the mixing of the signals of two Analog-to-digital converter (ADC)'s I/Q channels and one MS at about 6.6-GHz frequency to output a readout pulse. Both the employed DAC and ADC are field-programmable-gate-array-controlled and respond at the nanosecond scale. The output from the circuit, before being captured and demodulated by the ADC, is sequentially amplified by an impedance-transformed Josephson parametric amplifier (JPA, which is pumped by a 13.5-GHz MS and modulated by a DC bias), a high electron mobility transistor (HEMT), and two room temperature amplifiers. Furthermore, a few custom-made circulators, attenuators\textcolor{black}{, and} filters are utilized at some specific locations of the signal lines to reduce the noise that may affect the operations of the device.}
	\label{Schematic lay out of the experimental setup}
\end{figure*}

Our device consists of five frequency-tunable superconducting Xmon qubits\textcolor{black}{, labeled as $Q_{j}$ ($j=$1 to 5)}, each with an anharmonicity of approximately $2\pi\times240$ MHz. Every Xmon qubit has a microwave line (XY line) to drive its state transitions and an individual flux line (Z line) to dynamically tune its frequency. These two constituents consequently make each qubit flexibly on-and-off coupled (with a coupling strength $g_{b,j}$) to \textcolor{black}{a bus resonator} $R_b$ with a bare frequency $\omega_{b}/2\pi \simeq \SI{5.582} {\giga\hertz}$ and an energy relaxation time $T_b \simeq 13$ $\mu$s. Besides, each qubit is also dispersively coupled to its own readout resonator, whose frequency and leakage rate are denoted by \textcolor{black}{$\omega_{r,j}$ and $\kappa_{f,j}$,} respectively. All the readout resonators are coupled to a common transmission line to enable the multiplexed readout of all qubits' states. It is worth pointing out that the readout measurement performed here features both single-shot and quantum nondestructive characteristics, and is achieved with the assistance of an impedance-transformed Josephson parametric amplifier (JPA) with a bandwidth of about $150$ MHz. In this experiment, \textcolor{black}{the NH Hamiltonian dynamics is realized by coupling $Q_1$ to its readout resonator $R_1$,  and $Q_2$ serves as an ancilla qubit for reading out the joint $Q_1$-$R_1$ output state. The parameters of $Q_j$ and $R_j$ ($j=1,2$) are} listed in TABLE~\ref{table1}, \textcolor{black}{including} energy relaxation time $T_{1,j}$, Ramsey Gaussian dephasing time $T_{2,j}^{\star}$, and spin echo Gaussian dephasing time $T_{2,j}^{SE}$ at their idle frequency $\omega_{id,j}$. \textcolor{black}{The readout fidelity ($F_{k,j}$) is defined as the probability of correctly reading out the state of $Q_j$ when it is in $|k\rangle$. For simplicity, we will omit the subscript ``1''  of the test qubit and its readout resonator, and use the subscript ``a''  to denote the ancilla qubit.} The detailed experimental setup, including the whole electronics and wiring for the device control, is summarized in Fig.~\ref{Schematic lay out of the experimental setup}. The readout resonator for $Q$ is painted blue as an emphasis in the figure, as it is also used as a decaying resonator for constructing the desired non-Hermitian dynamics.

\begin{table*}[hb]
	\centering
	\renewcommand{\arraystretch}{1.9}
	\begin{tabular}{{ccc}}
		\hline\hline
		\centering
		Parameters &  $Q_1$ ($Q$) & \qquad\qquad\qquad\qquad  $Q_2$ ($Q_{a}$)  \qquad\qquad\qquad\qquad \qquad   \\ \hline
		\centering
		Qubit idle frequency, $\omega_{id,j}/2\pi$ &      5.99 GHz            & 5.23 GHz  \\
		Coupling strength to the bus resonator $R_b$, $g_{b,j}/2\pi$& 20.9 MHz & 20.3 MHz \\
		Coupling strength to the decaying resonator $R_j$, $g_{r,j}/2\pi$&  41 MHz & 40 MHz\\
		Energy relaxation time, $T_{1,j}$ &    14.3 $\mu$s              &        24.8 $\mu$s        \\
		Ramsey dephasing time, $T^{\star}_{2,j}$  &      5.3 $\mu$s           &  1.1 $\mu$s\\
		Dephasing time with spin echo, $T^{SE}_{2,j}$ &    14.7 $\mu$s       & 3.5 $\mu$s \\
		Frequency of decaying resonator, $\omega_{r,j}/2\pi$ &  6.66 GHz   &  6.76 GHz\\
		Leakage rate of decaying resonator, $\kappa_{f,j}$ &   1/200 $ns^{-1}$  & 1/226 $ns^{-1}$  \\
		$|g \rangle$ state readout fidelity, $F_{g,j}$ &      0.981       &  0.977 \\
		$|e \rangle$ state readout fidelity, $F_{e,j}$&       0.901       &  0.902 \\ \hline\hline
	\end{tabular}
	\caption{\label{table1} \textbf{Parameters of the circuit QED system.} \textcolor{black}{The parameters of both the test qubit ($Q_1$) and ancilla qubit ($Q_2$) are measured at their idle frequencies $\omega_{id,j}$ ($j=1,2$).} $\omega_{id,j}$ is also the point at which the single-qubit rotations and state tomographies are performed. \textcolor{black}{$g_{b,j}$ denotes the $R_b$-$Q_j$ coupling strength, which was inferred from the quantum Rabi signals of $Q_j$ resonantly coupled to $R_{b}$. $g_{r,j}$ is the $Q_{j}$-$R_{j}$ coupling strength, which was} deducted by measuring the dispersive frequency shift of the decaying resonator. The fidelity for correctly recording each qubit's state in experiment is $F_{k,j}$, characterized by extracting the state information of each readout resonator with the resonance frequency and leakage rate $\omega_{r,j}$ and $\kappa_{f,j}$, respectively.} 
	%\caption{\label{table1} \textbf{Qubits characteristics.} The resonator energy relaxation time $T_r = \SI{9.6}{\micro\second}$. }
\end{table*}

\section{Controlled $Q$-$R$ sideband couplings}

\label{subsection1A}
In a superconducting circuit, the parametric modulation is often achieved by modulating the flux. In our work, the modulation protocol is implemented by \textcolor{black}{applying} an external flux of the form
\begin{equation}
\Phi_{ext}(t)=\overline{\Phi}+\tilde{\Phi}\cos{(\nu^{\prime}t)},
\end{equation}
to tune \textcolor{black}{the transition frequency of the superconducting qubit $Q$.} Here, $\overline{\Phi}$ is the parking flux, $\tilde{\Phi}$ and $\nu^{\prime}$ are, respectively, the modulation amplitude and frequency. \textcolor{black}{The Josephson energy of $Q$ is modified} through an external flux $\Phi_{ext}$,
\begin{equation}
\label{EJ}
E_J(t)=E_{J\sum}\left\vert\cos\left[\pi\frac{\Phi_{ext}(t)}{\Phi_0}\right]\right\vert,
\end{equation}
with $\Phi_0=h/(2e)$ being the flux quantum. Under \textcolor{black}{this modification, the transition frequency of $Q$ is modulated as (setting $\hbar =1$)} 
\begin{equation}
\label{we}
 \omega_e(t)\simeq \sqrt{8E_cE_J(t)}-E_c,
\end{equation}
where $E_c$ represents the charge energy. In light of the sinusoidal function in $E_J(t)$ of Eq.~(\ref{EJ}), Eq.~(\ref{we}) can be evaluated by Fourier series expansion, that is,
\begin{equation}
\label{we_t}
\omega_e(t)=\omega_0+\sum\limits_{k=1}^{\infty}\varepsilon_k\cos{(\omega_kt)}.
\end{equation}
Here, $\omega_0$ stands for the averaged $Q$ transition frequency, and $\varepsilon_k$ and $\omega_k$ denote the amplitude and frequency of the $k$-th harmonic, respectively.

Due to the nonlinear flux dependence of $\omega_e(\Phi_{ext})$, the average value $\omega_0$ of the $Q$ transition frequency will be shifted away from $\omega_e(\overline{\Phi})$ by some amount, and this shifted frequency amount can be measured by a Ramsey interferometer. In this experiment, the operation point during the modulation procedure is chosen at the sweet point of $Q$ with $\Phi_{ext}=0$ and $\overline{\Phi}=0$. This is because at the $Q$'s sweet point, Eq.~(\ref{we_t}) can be substantially simplified to the following compact expression,
\begin{equation}
\label{feapprox1}
\omega_e(t)\approx \omega_0+\varepsilon\cos{(\nu t)},
\end{equation}
by simply keeping one dominant Fourier component while neglecting all the rest higher-order harmonic terms. Here the actual qubit modulation frequency $\nu$ is twice that of the corresponding flux modulation \cite{4}. In the experiment, $\varepsilon$ can be readily manipulated by tailoring the \textcolor{red}{z-pulse amplitude (zpa)}. 

\textcolor{black}{Under this parametric modulation, the coherent dynamics of the system combined by $Q$ and $R$ is governed by the Hamiltonian (setting $\hbar =1$)}
\begin{equation}
\label{H}
H = \omega_e(t)|e\rangle \langle e|+ \omega_r a^\dagger a
+ g_r(a^\dagger |g\rangle \langle e|+a|e\rangle \langle g|),
\end{equation}
where $\omega_r$ is the center frequency of the quantized decaying bosonic mode and $g_r$ ascribes \textcolor{black}{the on-resonance coupling strength between $Q$ and $R$.} Now, substituting Eq.~(\ref{feapprox1}) into Eq.~(\ref{H}) and working in the interaction picture would transform the full Hamiltonian of Eq.~(\ref{H}) into
\begin{equation}
H_I = g_r e^{i\Delta_rt-i\mu\sin(\nu t)}a^{\dagger}|g\rangle \langle e|+H.c.,
\label{Hfull}
\end{equation}
where $\mu=\varepsilon /\nu$, $\Delta_{r}=\omega_{r}-\omega_0$, and $H.c.$ means the Hermitian conjugate. Using the Jacobi-Anger expansion
\begin{equation}
e^{i\mu\sin{\theta}}=\sum\limits_{-\infty}^{\infty}J_n(\mu)e^{in\theta}, \label{JA}
\end{equation}
with $J_n(x)$ being the $n$-th Bessel function of the first kind, Eq.~(\ref{Hfull}) then becomes
\begin{equation}
\begin{split}
H_I &= g_r\left[\sum\limits_{n=-\infty}^{\infty}J_n(\mu)e^{-i(n\nu-\Delta_r)t}a^{\dagger}|g\rangle \langle e|+H.c.\right].
\end{split}
\label{HI}
\end{equation}

%From Eq.~(\ref{HI}), we notice that when the modulation frequency satisfies $\nu=\Delta_r $ $(\nu=\frac{1}{2}\Delta_r)$ which corresponds to the first(second)-order sideband modulation for the coupling between $Q$ and $R$, a swap $|e,0_{r}\rangle \leftrightarrow |g,1_{r}\rangle$ is available. In this way, we arrange our modulation protocol and confine the system’s energy-level structure to the configuration shown in Fig.~1 in the main text.

Equation~(\ref{HI}) looks complicated and time-dependent. In practice, its complexity and time-dependence can be easily removed \textcolor{black}{under the conditions $\nu=\Delta_{r}$ ($\nu=\Delta_{r}/2$) and $g_{r}\ll \nu$.} Accordingly, this frequency setting is referred to as the first-order (or second-order) sideband modulation for establishing the $Q$-$R$ coupling with the coupling strength $\Omega=J_1(\mu)g_r$ (or $\Omega=J_2(\mu)g_r$). As a consequence, a swap operation of $|e,0\rangle \leftrightarrow |g,1\rangle$ is available under such a modulation arrangement.

\begin{figure*}[htbp] 
	\centering
	\includegraphics[width=7in]{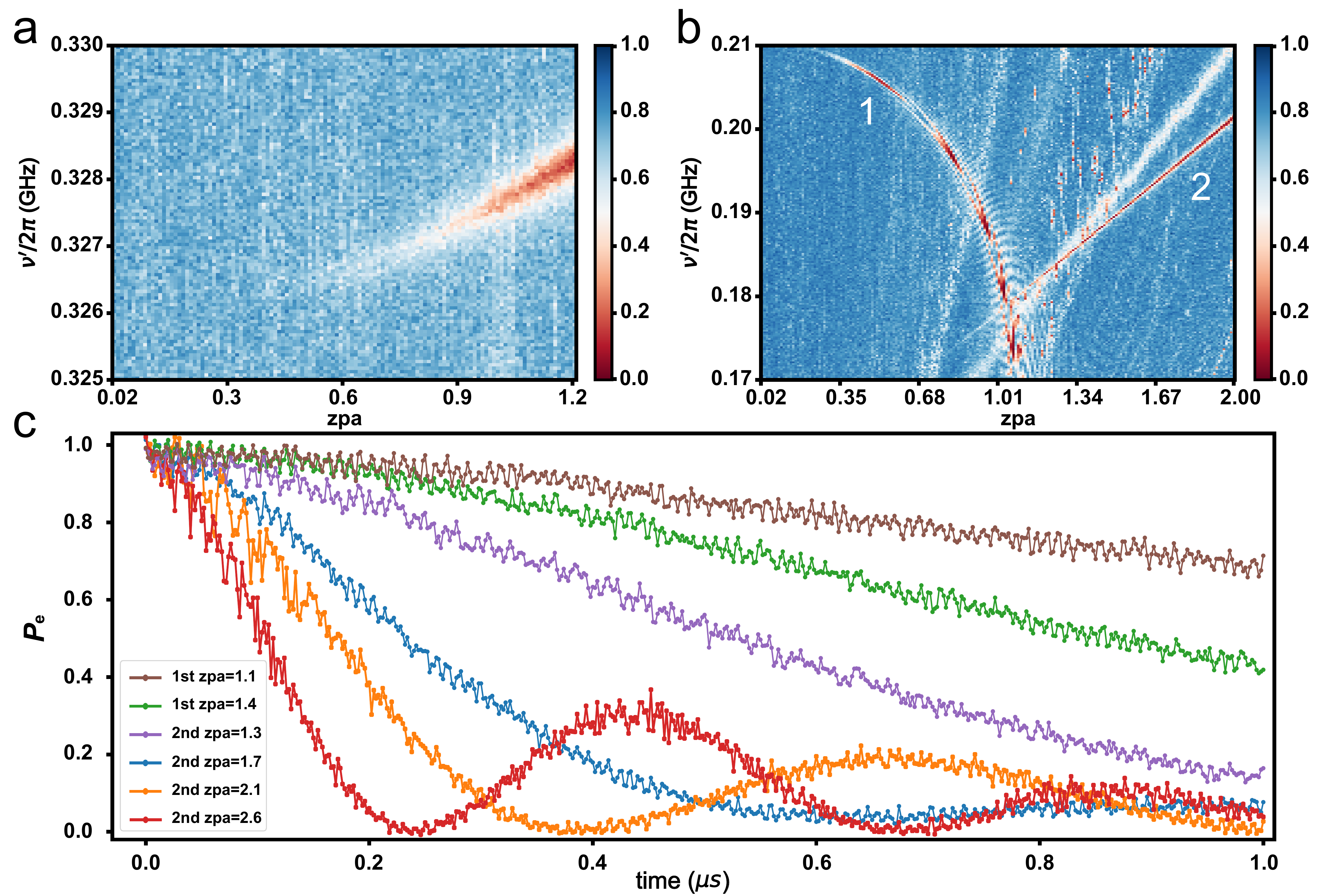}
	\caption{\textbf{Scanning map of the population $P_e$ of the superconducting qubit $Q$ after 1-$\mu s$ parametric modulations.}
		(a) The measured data obtained by coupling $Q$ with $R$ via the first-order sideband modulation. (b) The measured data labeled as ``1'' (``2") attained by establishing the $Q$-$R_b$ ($Q$-$R$) interaction via the first-order (second-order) sideband modulation. Here, the $x$-axis states the modulation amplitude in zpa and the $y$-axis gives the modulation frequency. (c) Temporal evolution of the population $P_e$ under different single-tone modulations.}
	\label{singleTone}
\end{figure*}

%\textcolor{black}{To optimize the Rabi signals, we use different sideband couplings at different regimes. When ?, the test qubit is coupled to its resonator resonator at the first sideband with respect to the parametric modulation, produced by an ac flux with a modulating frequency of ?. For ?, the coupling is induced at the second sideband of an parametric modulation with a modulating frequency of ?. To confirm the validity of these modulations, in FigS. ? and ? we present the vaccum Rabi oscillations signals for the test qubit induced by these sideband couplings for different modulating amplitudes.}
\textcolor{black}{Without considering the interference of $R_{b}$, it is favorable to use the second-order sideband coupling due to the limitation of the available modulating flux. However, when this interaction is weak, it may be strongly intervened by the first-order sideband coupling associated with $R_{b}$. To optimize the parametric modulation, we prepare $Q$ in the $|e\rangle$-state, and observe the output $|e\rangle$-state population ($P_{e}$) after a modulating pulse with a fixed duration of 1 $\mu s$. Fig.~\ref{singleTone}a and Fig.~\ref{singleTone}b display the populations, measured respectively under the first-order and second-order sideband modulations.}
\textcolor{black}{
Figure ~\ref{singleTone}b shows a crossing region where the qubit $Q$ \textcolor{black}{is effectively coupled to} both resonators $R$ and $R_b$. In this region, the first-order sideband interaction \textcolor{black}{(labeled with ``1")} with $R_b$ coincides with the second-order sideband interaction \textcolor{black}{(labeled with ``2")} with $R$, resulting in undesired effects in our experiments.}
\textcolor{black}{To circumvent this issue, we use the first-order sideband modulation to realize $Q$-$R$ swapping interaction for the corresponding region.}
\textcolor{black}{
 To confirm the validity of these modulations, in Fig.~\ref{singleTone}c we present the vacuum Rabi oscillation signals for the test qubit induced by these sideband couplings for different modulating amplitudes.
}
In this way, we arrange our modulation protocol and confine the system’s energy-level structure to the configuration shown in Fig.~1 of the main text.

\section{Numerical simulations of the non-Hermitian Hamiltonian dynamics}
%To ensure that the system dynamics can be well described by the effective Hamiltonian $\cal{H}_{\textbf{NH}}$ in the main text, we have carried out similar calculations using the full Hamiltonian $H_I$ of Eq.~(\ref{Hfull}) given above. In Fig.~\ref{num}, we have presented some representative numerical results calculated by these two Hamiltonians for comparison. In these calculations, we have uniformly utilized the Lindblad master equation to compute the temporal evolution of our system (setting $\hbar =1$),
%\begin{equation}
%\frac{\partial\rho}{\partial t}=-i[H_I,\rho]+\sum\limits_{k=q,r}\left(L_k\rho L_k^{\dagger}-\frac{1}{2}\{L_k^{\dagger}L_k,\rho\}\right),
%\label{master}
%\end{equation}
%with the Lindblad dissipators $L_q=\sqrt{\kappa_{q}}\sigma^{-}$ and $L_r=\sqrt{\kappa_{f}}a$, the decay rates $\kappa_{q}$ and $\kappa_f$ of the qubit and microwave field, and $\sigma^-$ the spin transition operator. 
%\textcolor{black}{
%In order to study the behavior of the system within the subspace $\{|g,1\rangle, |e,0\rangle\}$, we exploit the post-selection method to effectively eliminate any dynamics outside this subspace. This approach allows us to describe the system's behavior within this subspace using an effective non-Hermitian Hamiltonian $\cal{H}_{\textbf{NH}}$. }

\begin{figure*}[htbp] 
	\centering
	\includegraphics[width=6in]{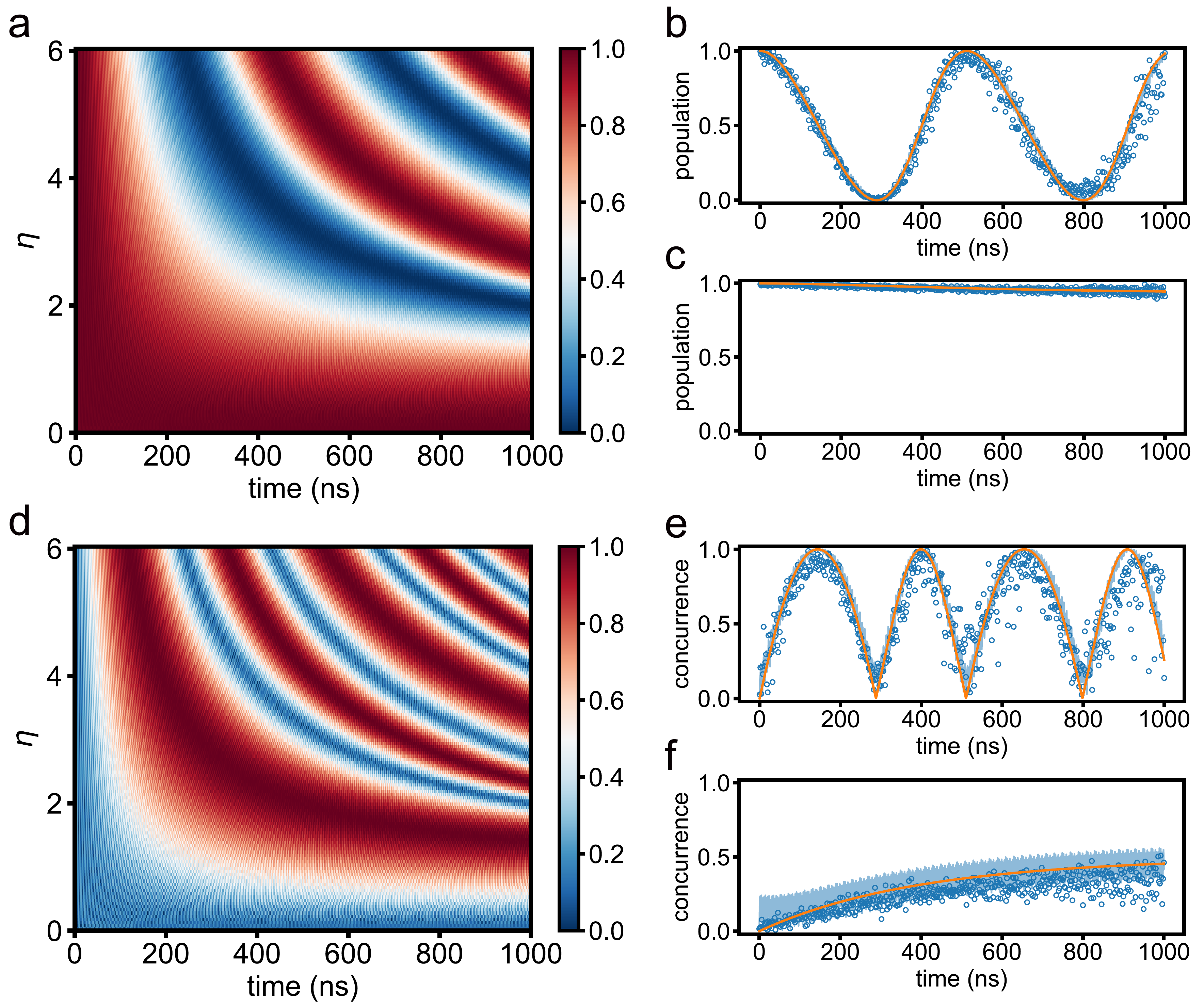}
	\caption{\textbf{Numerical evolutions of the system's non-Hermitian evolution using the full Hamiltonian $H_I$.} (a) Vacuum Rabi oscillations by calculating $P^N_{|e,0\rangle}$ in terms of the rescaled coupling $\eta=4\Omega/\kappa$. As an example, (b) and (c) compare $ P^N_{|e,0\rangle}$ obtained respectively with use of the full (blue) and effective (orange) Hamiltonians before ($\eta =5$) and after ($\eta=0.5$) the exceptional entanglement transition. (d) Concurrence $\cal{E}$ evolution for different $\eta $. As an example, (e) and (f) compare $\cal{E}$ obtained respectively with use of the full (blue) and effective (orange) Hamiltonians before ($\eta =5$) and after ($\eta=0.5$) the exceptional entanglement transition. The blue empty circles are experimental data.}
	\label{num}
\end{figure*}

\textcolor{red}{To validate the authenticity of the effective NH Hamiltonian presented in Eq. (1) of the main text, we conducted a numerical simulation focusing on the system dynamics associated with the no-jump trajectory, which is governed by the original NH Hamiltonian. This NH Hamiltonian integrates the original coherent Hamiltonian from Eq. (\ref{Hfull}) with the NH terms, namely the last two terms of Eq. (1) in the main text. Figures \ref{num}a to f illustrate the evolution of the population $|e,0\rangle$ and the qubit-resonator concurrence corresponding to the no-jump trajectory, determined by the original NH Hamiltonian. These results are juxtaposed with those obtained using the effective NH Hamiltonian from Eq. (1) in the main text. The comparative analysis confirms a robust correspondence between the dynamics predicted by the effective NH Hamiltonian and those governed by the original NH Hamiltonian.}
%
%\textcolor{black}{To confirm the validity of the approximations for obtaining the effective Hamiltonian, we perform numerical simulations with the full Hamiltonian. The results are presented in Figs.~\ref{num}a to f.}
%It is worth noting that the evolution of the entanglement differs significantly from the population dynamics. 
%
In addition to the doubled oscillation frequency, the entanglement is much more sensitive to the control parameter. For instance, when the population of state $|g,1\rangle$ changes from $0$ to $0.1$, the concurrence increases by an amount $\sim0.6$, which is one order of magnitude larger than the variation in the population of state $|g,1\rangle$. Therefore, \textcolor{black}{the concurrence (Figs.~\ref{num}e and f) is much more influenced by the high-frequency oscillating terms than the population (Fig.~\ref{num}b and c). In these figures, the orange and blue lines respectively denote the numerical results with the effective and full Hamiltonians, and the circles are the experimental data.}

\begin{figure*}[htbp] 
	\centering
	\includegraphics[width=5in]{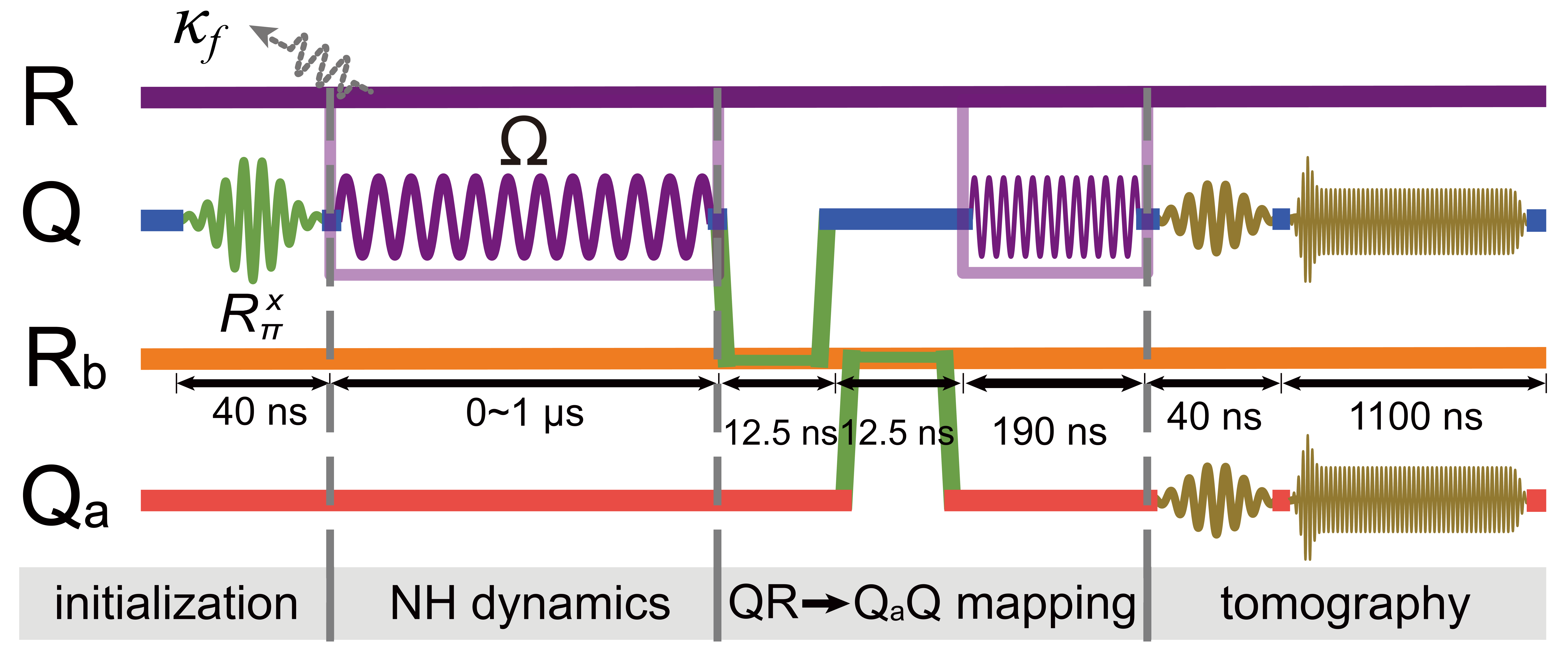}
	\caption{\textbf{Pulse sequence.}  
	 The NH dynamics starts with the initial state $\left\vert e,0\right\rangle $, prepared from $\left\vert g,0\right\rangle $ with a $\pi$ pulse. After a preset evolution time, the parametric modulation is switched off, followed by the state mappings $Q\rightarrow R_{b}$, $R_b\rightarrow Q_a$, and $R\rightarrow Q$, each realized by an on-resonance swapping gate, where $Q_{a}$ is an ancilla qubit, and $R_{b}$ represents the bus resonator coupled to both qubits. After these mappings, the $Q$-$R$ output state produced by the NH dynamics is encoded in the joint $Q_{a}$-$Q$ state, which is then measured by quantum state tomography.}
	\label{pulse sequence}
\end{figure*}

\section{Joint $Q$-$R$ quantum state tomography}

The joint $Q$-$R$ output state after the NH dynamics is read out by mapping it to the $Q_{a}$-$Q$ system.
\textcolor{blue}{The pulse sequence is displayed in Fig.~\ref{pulse sequence}.}
To accomplish this mapping, we must first transfer the state of $Q$ to $R_b$. This is realized by tuning $Q$ on resonance with $R_b$ for a duration $t=\pi/(2g_{b})=12.5$ ns. After the $Q\rightarrow R_b$ state transfer, we reset $Q$ to its idle frequency and bring $Q_a$ resonance with $R_b$. Similarly, after an interaction duration $t$, $Q_a$ carries the bus resonator's state. Up to this moment, we complete the state transfer from $Q$ to $Q_a$.

The next step is to transfer the $R$-state to $Q$. Because the $R$-frequency  is much higher than the $Q$'s, it is impossible to realize a quantum state transfer within the same short time period using the resonant coupling technique introduced above. As such, we consider the second-order sideband resonant coupling proposed in Section~\ref{subsection1A} instead. Limited by the experimental constraints, the maximum achievable coupling strength is about $\Omega_{max} = 2\pi \times 1.18$ MHz, which gives rise to the fastest state transfer time duration of approximately $\tau=193$ ns. This period is comparable to the lifetime of the readout resonator 
\textcolor{black}{so that the dissipation of $R$ has a significant impact on the state mapping.}
	
\textcolor{black}{However, there is a one-to-one correspondence between the $Q_{a}$$Q$ output state after the state mapping and $Q$-$R$ state just before the mapping for the no-jump case, as interpreted below. Without loss of the generality, the $Q$-$R$ state right before the state mapping can be expressed as
\begin{equation}
|\psi_0\rangle = c_1 |e,0\rangle +c_2|g,1\rangle,
\end{equation}
with $|c_1|^2+|c_2|^2=1$. After the state mapping, the $Q_{a}$-$Q$ output state is
\begin{equation}
|\psi_1\rangle = (1/\sqrt{|c_1|^2+k^2|c_2|^2})(c_1|e_a,g\rangle+kc_2|g_a,e\rangle),
\end{equation}
where
%$k=(\Omega_{max}/\sqrt{\Omega_{max}^2-\kappa^2/16})e^{-\kappa t}e^{-\kappa \tau/4}\sin \left( \sqrt{\Omega _{\max }^{2}-\kappa^{2}/16}\tau \right)$.
$k=e^{-\kappa_f t}e^{-\kappa_f \tau/4}$.
For simplicity, we here do not include the phase accumulated during the mapping. This implies that the original $Q$-$R$ output state $|\psi_0\rangle$ can be inferred from $|\psi_1\rangle$ by multiplying the coefficient of the component $|g_{a},e\rangle$ by $1/k$ and then renormalizing the resulting state. In the experiment, the $Q_{a}$-$Q$ output state is characterized by the two-qubit density matrix, which is reconstructed through joint quantum state tomography. The result associated with the no-jump trajectory is obtained by projecting the density matrix to the single-excitation subspace $\{|e_{a},g\rangle,|g_{a},e\rangle\}$, which can be expressed as
\begin{equation}
\rho_1 = 
\begin{pmatrix}
\rho_{11} & \rho_{12} \\
\rho_{21} & \rho_{22} \\
\end{pmatrix}
\end{equation}
The elements of $Q$-$R$ density matrix within $\{|e,0\rangle,|g,1\rangle\}$ right before the state mapping are related to those of $\rho_1$ by
\begin{equation}
\rho_1 = \frac{1}{\rho_{11}+\rho_{22}/k^2}
\begin{pmatrix}
\rho_{11} & \rho_{12}/k \\
\rho_{21}/k & \rho_{22}/k^2 \\
\end{pmatrix}
\end{equation}	
}

\section{Qubit readout corrections}
\label{Qubit readout corrections}
\begin{figure*}[htbp] 
	\centering
	\includegraphics[width=6in]{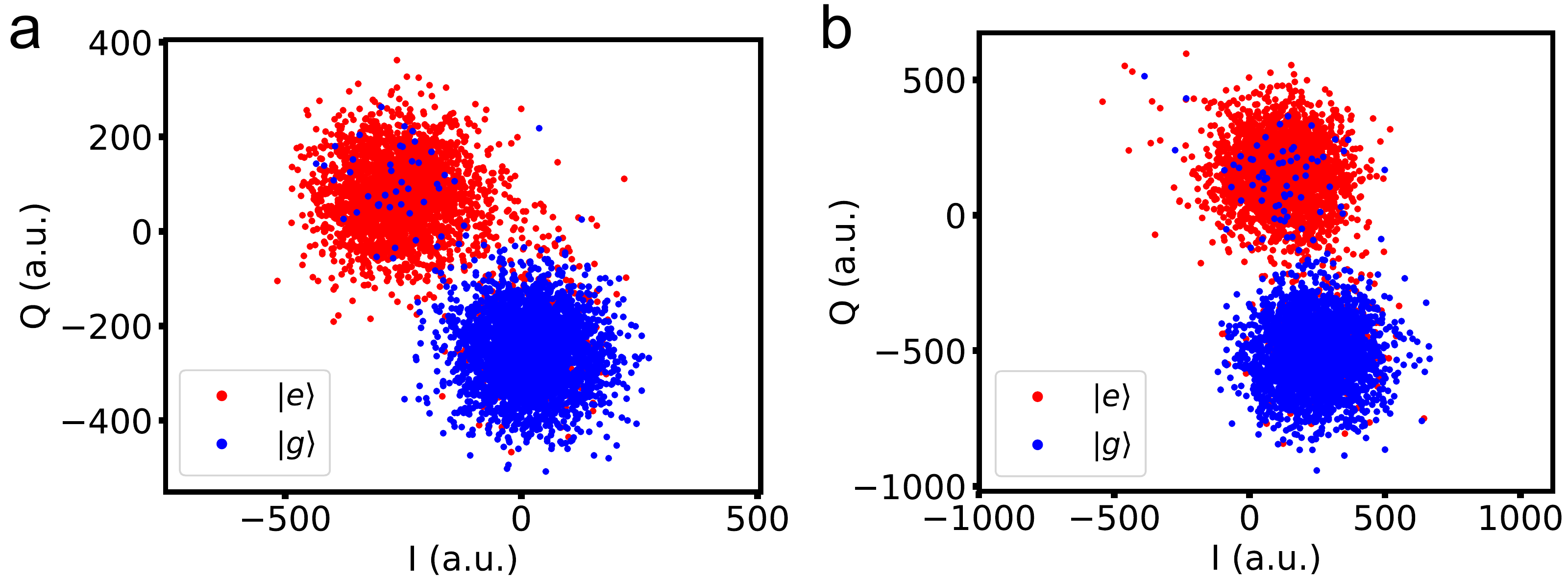}
	\caption{\textbf{Qubit readouts for} $Q$ \textbf{(a)} and $Q_a$ \textbf{(b)} with 3000 repetitions and 1.1-$\mu s$ readout duration. }
	\label{IQplot}
\end{figure*}

The fidelity matrix for calibrating the measured probabilities is defined as 
\begin{equation}
\hat{F} = 
\begin{pmatrix}
F_{g} & e_{ge} \\ 
e_{eg} & F_{e} \\
\end{pmatrix},
\label{fidelity}
\end{equation}
where $F_{j}$ ($j=g,e$) represents the probability \textcolor{black}{for correctly reading out the state of the qubit when it is in $|k\rangle$, and $e_{jk}$} ($j,k=g,e$) stands for the error that describes the leakage probability from the state $|k\rangle$ to state $|j\rangle$. \textcolor{black}{To illustrate how the qubit readout error can be corrected, we denote the measured probability distribution as $\hat{P}_{M}$ and the genuine probability distribution as $\hat{P}_{N}$. The relation among} $\hat{F}$, $\hat{P}_{M}$, and $\hat{P}_{N}$ is established by the following simple identity,
\begin{equation}
\hat{P}_{M} = \hat{F} \cdot \hat{P}_{N}. \label{relation}
\end{equation}
The above relation (\ref{relation}) implies that the genuine states of the system can be mathematically reconstructed by performing the matrix inversion of $\hat{F}$, i.e., $\hat{P}_N=\hat{F}^{-1}\cdot\hat{P}_M$. The data used in our calibrations are extracted from the measured $I$-$Q$ (in phase and quadrature) values, as shown in Fig.~\ref{IQplot}.
According to the measured data, in this work the fidelity matrices for $Q$ and $Q_a$ are   
\begin{equation}
\hat{F}_{Q} \simeq 
\begin{pmatrix}
0.981 & 0.099 \\ 
0.019 & 0.901 \\
\end{pmatrix}, \quad
\hat{F}_{Q_a} \simeq 
\begin{pmatrix}
0.977 & 0.098 \\ 
0.023 & 0.902 \\
\end{pmatrix}.
\end{equation}
In the computational basis, the joint QST measurement on $Q$ and $Q_a$ can be therefore corrected by the following formula,
\begin{equation}
\hat{P}_{N} = (\hat{F}_{Q}\otimes\hat{F}_{Q_a})^{-1}  \cdot \hat{P}_{M}.
\end{equation}

%Due to the long duration of the experiment, the fidelity matrices obtained from the measurements may in practice differ from moment to moment. Even if the same fidelity matrix is used for all data processing, there is still some readout error. To characterize this error, we have used 20 different fidelity matrices randomly measured between individual experimental runs for calibration, and have accordingly attained 20 sets of data in this work. The error bars are the standard deviation of these 20 sets of data.

\section{Extraction of eigenenergies and eigenstates of the non-Hermitian Hamiltonian}
After collecting the data, we have applied the least-squares fitting to the measured density matrix $\rho$ in order to extract the eigenenergies and eigenstates of the system. Theoretically, for a certain point in the parameter space, we can in principle write the eigenenergies and the corresponding eigenstates of the non-Hermitian system as
%\begin{equation}
%\begin{split}
%E_+ = c_{+,1}  + ic_{+,2} |e,0\rangle , \quad |\Phi_+\rangle = \alpha_{+} |g,1\rangle + \beta_{+} |e,0\rangle \\
%E_- = c_{-,1}  + ic_{-,2} |e,0\rangle , \quad |\Phi_-\rangle = \alpha_{-} |g,1\rangle + \beta_{-} |e,0\rangle,
%\end{split}
%\end{equation}
\begin{equation}
\begin{split}
E_{\pm} = c_{\pm,1}  + ic_{\pm,2} , \quad |\Phi_{\pm}\rangle = \alpha_{\pm} |g,1\rangle + \beta_{\pm} |e,0\rangle ,
\end{split}
\end{equation}
%where $\alpha_{\pm} = \alpha_{\pm,1} + i \alpha_{\pm,2}$, $\beta_{\pm} = \beta_{\pm,1} + i \beta_{\pm,2}$, and the undetermined parameters ($c_{\pm,1}$, $c_{\pm,2}$, $\alpha_{\pm,1}$, $\alpha_{\pm,2}$, $\beta_{\pm,1}$ and $\beta_{\pm,2}$) are real numbers. Assuming the system's initial state is
%
where $c_{\pm,1}$, $c_{\pm,2}$, $\alpha_{\pm}$ and $\beta_{\pm}$ are the fitting parameters. If we assume the system's initial state to be
\begin{equation}
|e,0\rangle = \frac{1}{\alpha_-\beta_+-\alpha_+\beta_-}(\alpha_- |\Phi_+\rangle - \alpha_+ |\Phi_-\rangle),
\end{equation} 
then at time $t$ the system will evolve into the following state
\begin{equation}
|\psi(t)\rangle = \frac{1}{\alpha_-\beta_+-\alpha_+\beta_-}(\alpha_- e^{-iE_+ t} |\Phi_+\rangle + \alpha_+  e^{-iE_- t} |\Phi_-\rangle).
\end{equation} 
To minimize the fitting errors, we have defined an error function as a guidance, which has the form of 
\begin{equation}
\mathrm{erf} = \mathrm{Tr}[\rho|\psi(t)\rangle \langle\psi(t)|] -1.
\label{erf}
\end{equation}
Based on this error function, the next step is to seek the fitting parameters which can minimize Eq.~(\ref{erf}) at each moment. In this way, a reliable least-squares fitting to the entire evolution of the density matrix $\rho$ is found with all the parameters: $c_{\pm,1}$, $c_{\pm,2}$, $\alpha_{\pm}$, $\beta_{\pm}$.

\begin{figure*}[h] 
	\centering
	\includegraphics[width=6in]{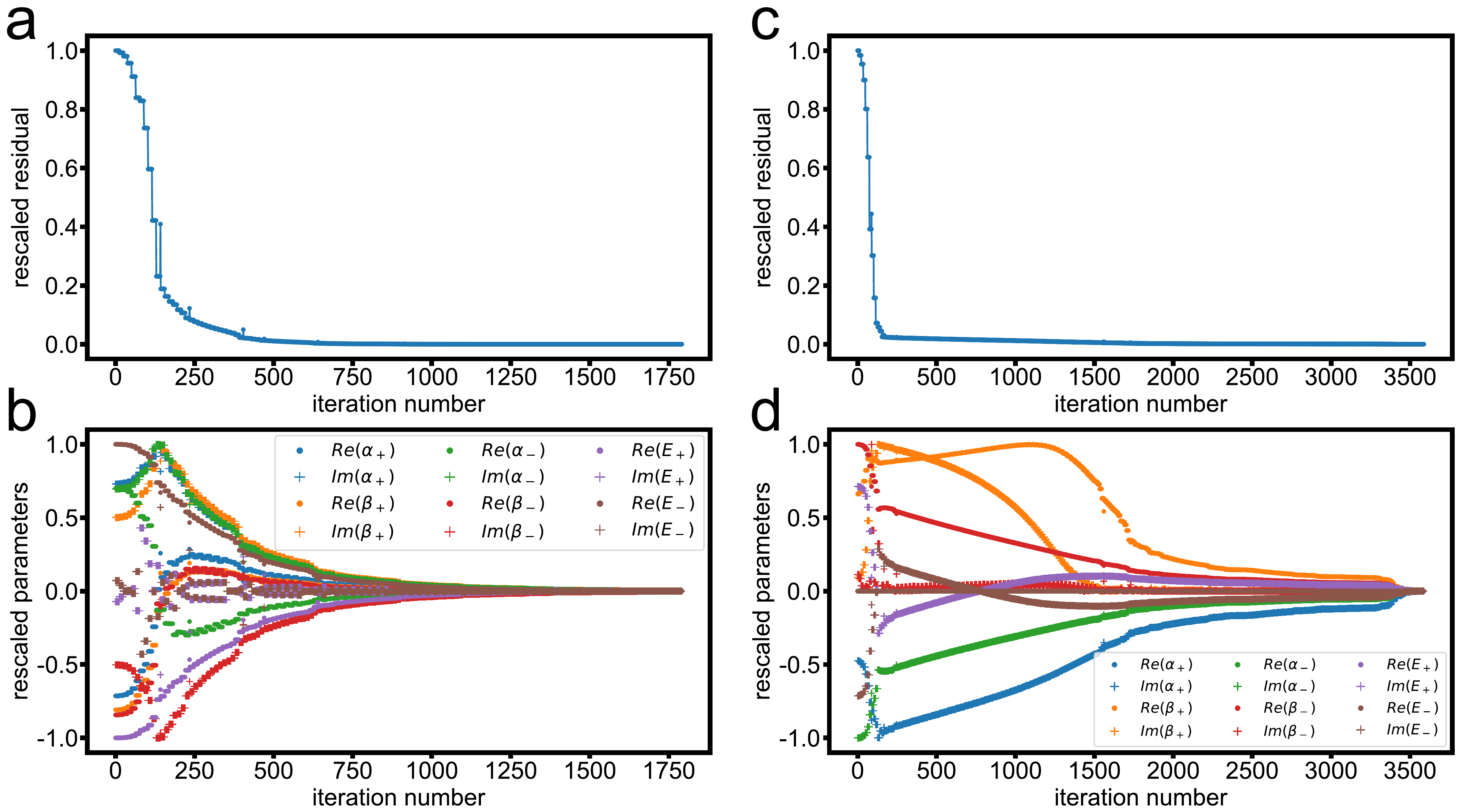}
	\caption{\textbf{Convergence of residuals and parameters.} (a) (b) Above the EP, $\eta=5$. (c) (d) Below the EP, $\eta=0.5$.}
	\label{fit_convergence}
\end{figure*}

\textcolor{black}{The convergence condition is critical for ensuring the success of a fitting process. In Fig.~\ref{fit_convergence}, we show the convergence of residuals and all parameters as \textcolor{black}{functions} of the number of iterations. To facilitate comparison, we rescaled each parameter to an appropriate size, using $[f_i-f_N]/[max(f)-min(f)]$, where $f$ represents the set of results after each iteration, $f_i$ represents the result of the $i$-th iterations, and $i=0,1,2,...,N$, \textcolor{black}{with $N$ being} the maximum iteration number. We observe that all parameters converge together after a sufficient number of iterations. We provide two representative examples, one for $\eta=5$ (above the EP, shown in Figs.~\ref{fit_convergence}a and b) and the other for $\eta=0.5$ (below the EP, shown in Figs.~\ref{fit_convergence}c and d). We verified the validity of the remaining fitting procedures using the same approach.}

%\textcolor{black}{To address concerns regarding fitting error, we have conducted a sensitivity analysis of our fitting procedure. In Fig. 3 of the main text, we displayed the fitting results using the corresponding ideal values as the initial inputs. To further demonstrate the impact of different initial guess values, we selected 10 additional sets of initial inputs with values ranging from 0.9$p_0$ to 1.1$p_0$, spaced 0.2$p_0$ apart, where $p_0$ represents the set of ideal values. The resulting average and standard deviation of the fitting results are shown in Fig. ?. We found that the standard deviation is small, which provides further evidence of the validity of our fitting procedure.}

\begin{figure*}[htbp] 
	\centering
	\includegraphics[width=5in]{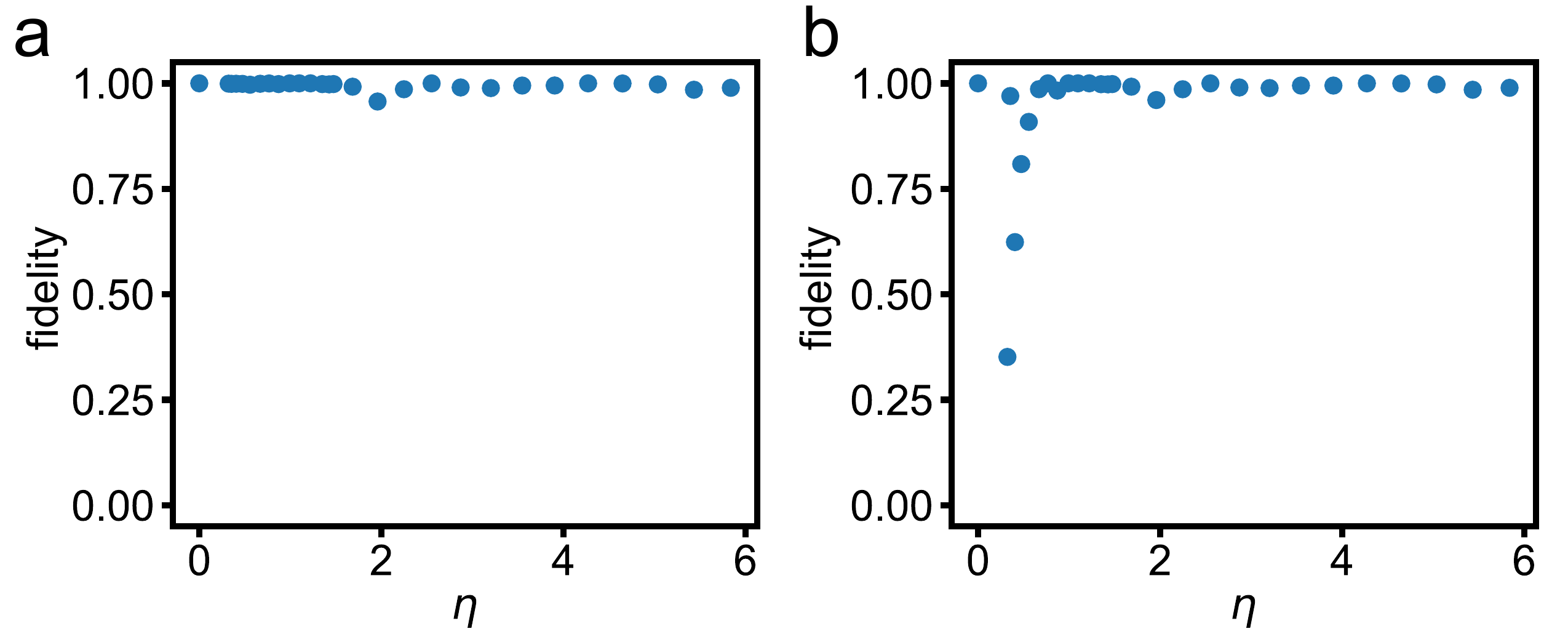}
	\caption{\textbf{Fidelities of fitted eigenstates.} (a) $\mathcal{F}_+$ as a function of $\eta$. (b) $\mathcal{F}_-$ as a function of $\eta$. }
\label{eigen_fid}
\end{figure*}

\begin{figure*}[h] 
	\centering
	\includegraphics[width=5in]{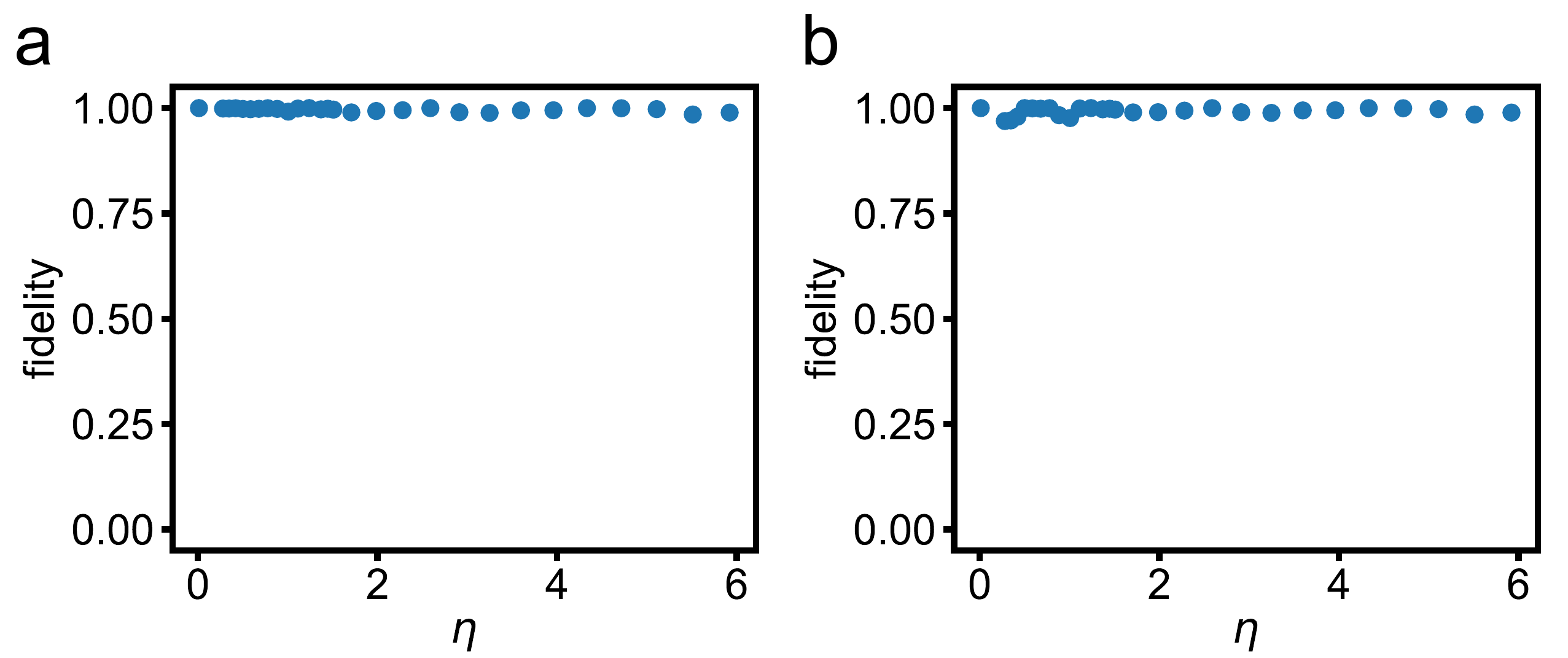}
	\caption{\textbf{Fidelities of fitted eigenstates after adjusting initial guesses.} (a) $\mathcal{F}_+$ as a function of $\eta$. (b) $\mathcal{F}_-$ as a function of $\eta$. }
	\label{eigen_fid_nodip}
\end{figure*}

We also calculate the fidelities of the eigenstates ($| \Phi_{\pm }^{^{\prime }}\rangle $) obtained in this manner with respect to the ideal ones $| \Phi _{\pm }\rangle $, which are defined as
 \begin{equation}
 {\cal F}_{\pm }=|\langle \Phi _{\pm }^{^{\prime }}| \Phi _{\pm }\rangle | ^{2}. 
 \end{equation}
The fidelities as functions of $\eta $ are presented in Figs.~\ref{eigen_fid}a and b, respectively. \textcolor{black}{The dip in Fig.~\ref{eigen_fid}b is mainly caused by two factors. First, we can observe from the comparison between Figs.~\ref{num}e and f that the effect of off-resonant terms in the parametric modulation is more significant when $\eta<1$. This is demonstrated by the oscillations in Fig.~\ref{num}f having a significantly larger amplitude than those in Fig.~\ref{num}e. Second, when $\eta<1$, the system's evolution is likewise more susceptible to dephasing, resulting in an overall measured concurrence (associated with the non-diagonal elements of the density matrix) that is lower than the theoretical value, as illustrated in Fig.~\ref{num}f. Therefore, the fitting error grows as the observed result deviates more from the theoretical value, leading to lower fidelity. Additionally, we discovered that when A<1, the fitting results are more sensitive to the initial guess and that adjusting the initial guess appropriately can lead to better fitting results, as shown in Fig.~\ref{eigen_fid_nodip}.} The results demonstrate that the eigenstates, extracted from the measured two-qubit output density matrices by our density-matrix post-projecting method, well agree with the ideal ones, associated with the no-jump evolution trajectories. This agreement confirms the validity of approximations for deriving the effective NH Hamiltonian, as well as the soundness of the density-matrix post-projecting method.

%
%\textcolor{black}{To confirm the validity of the procedure, we calculate the fidelities of thus-obtained eigenstates ($| \Phi_{\pm }^{^{\prime }}\rangle $) with respect to the ideal ones $| \Phi _{\pm }\rangle $, defined as \[{\cal F}_{\pm }=|\langle \Phi _{\pm }^{^{\prime }}| \Phi _{\pm }\rangle | ^{2}. \]The fidelities as functions of $\eta $ are presented in Fig. S?a and b, respectively. The results confirm that the extracted eigenstates well coincide with the ideal ones, which implies that the joint output state of the test qubit and its resonator associated with the no-jump evolution trajectory, measured by our density-matrix post-projecting method, is governed by a NH Hamiltonian that is in good agreement with the preset one.}

\section{Exceptional entanglement transition in a two-qubit system}

The exceptional entanglement transition is not restricted to the
light-matter system, but is a universal behavior for a variety of NH
interacting quantum systems. As a paradigmatic example, we here consider the
system \textcolor{black}{composed of} two decaying qubits interacting with each other by swapping
coupling \cite{1,2,3}. When
the two qubits have the same frequency, the no-jump evolution trajectory is
governed by the NH Hamiltonian (setting $\hbar =1$) 
\begin{equation}
{\cal H}_{NH}=\Omega (\sigma _{1}^{+}\sigma _{2}^{-}+\sigma _{1}^{-}\sigma
_{2}^{+})-\frac{i}{2}(\kappa _{1}\left\vert e_{1}\right\rangle \left\langle
e_{1}\right\vert +\kappa _{2}\left\vert e_{2}\right\rangle \left\langle
e_{2}\right\vert ), 
\end{equation}
where $\sigma _{j}^{+}=\left\vert e_{j}\right\rangle \left\langle
g_{j}\right\vert $ and $\sigma _{j}^{-}=\left\vert g_{j}\right\rangle
\left\langle e_{j}\right\vert $ with $\left\vert e_{j}\right\rangle $ ($%
\left\vert g_{j}\right\rangle $) denoting the upper (lower) level of the $j$th
qubit, $\kappa_{j}$ is the dissipation rate\ of $\left\vert
e_{j}\right\rangle $, and $\Omega $ is coupling strength. This swapping
coupling does not change the total excitation number of the system. When the
system is initially in a one-excitation state, its dynamics will be restricted
within the subspace $\{\left\vert e_{1},g_{2}\right\rangle ,\left\vert
g_{1},e_{2}\right\rangle \}$. In such a subspace, the eigenstates of the NH
Hamiltonian are given by
\begin{equation}
\left\vert \Phi _{\pm }\right\rangle ={\cal N}_{\pm }(\Omega \left\vert
e_{1},g_{2}\right\rangle +\Gamma _{\pm }\left\vert g_{1},e_{2}\right\rangle
), 
\end{equation}
where $ {\cal N}_{\pm }=(\Omega^2+|\Gamma_{\pm}|^2)^{-1/2}$ and $\Gamma _{\pm }=-i\kappa /4\pm E_{g}/2$ with $\kappa =\kappa
_{2}-\kappa _{1}$. The energy gap between these two eigenstates is $E_{g}=2%
\sqrt{\Omega ^{2}-\kappa ^{2}/16}$.

When the two qubits have distinct decaying rates, these eigenstates and
eigenenergies have the same forms as those of the qubit-resonator system.
Consequently, the energy gap undergoes a real-to-imaginary transition at the
EP $\eta =4\Omega /\left\vert \kappa \right\vert =1$, which is accompanied
by an entanglement transition of the eigenstates. The two-qubit concurrences \cite{5}
for the two eigenstates $\left\vert \Phi _{\pm }\right\rangle $ are 
\begin{eqnarray}
{\cal E}_{\pm }=\frac{2\Omega \left\vert \Gamma _{\pm }\right\vert }{%
	\left\vert \Gamma _{\pm }\right\vert ^{2}+\Omega ^{2}}. 
\end{eqnarray}
When $\eta \rightarrow 0$, the two eigenstates respectively reduce to $%
\left\vert e_{1},g_{2}\right\rangle $ and $\left\vert
g_{1},e_{2}\right\rangle $, each of which has no entanglement. When $\kappa_2 \ll \kappa_1$, the concurrence is increased linearly with $\eta$ until reaching the EP, where the energy gap vanishes and both eigenstates approximately converge to the same maximally entangled state
\begin{equation}
\left\vert \Phi _{\pm }\right\rangle =(\left\vert e_{1},g_{2}\right\rangle
-i\left\vert g_{1},e_{2}\right\rangle )/\sqrt{2}. 
\end{equation}
After crossing the EP, \textcolor{black}{$|\Phi_{\pm}\rangle$ move in opposite directions, but with the concurrences ${\cal E}_{\pm}$ remaining to be 1,}
independent of $\eta $.